\documentclass[iop]{emulateapj}
\usepackage{natbib}
\usepackage{amsmath}

\newcommand{\msol}{\,\textrm{M}_\sun}                

\shorttitle{The Dark Matter Dist. in Low-Mass Disk Galaxies: I}
\shortauthors{Relatores et al.}

\begin{document}

\title{The Dark Matter Distributions in Low-Mass Disk Galaxies. I. H$\alpha$ Observations Using the Palomar Cosmic Web Imager}
\author{Nicole C. Relatores$^{1,2}$, Andrew B. Newman$^{2,1}$, Joshua D. Simon$^2$, Richard Ellis$^3$, Phuongmai Truong$^4$, Leo Blitz$^4$, Alberto Bolatto$^5$, Christopher Martin$^6$, Patrick Morrissey$^6$}
\affil{$^1$ Department of Physics and Astronomy, University of Southern California, Los Angeles, CA, 90089-0484, USA\\
$^2$ Observatories of the Carnegie Institution for Science, Pasadena, CA 91101, USA\\
$^3$ Department of Physics and Astronomy, University College London, Gower Street, London WC1E 6BT, UK\\
$^4$ Department of Astronomy, University of California, Berkeley, CA 94720, USA\\
$^5$ Department of Astronomy, University of Maryland, College Park, MD 20742-2421, USA\\
$^6$Cahill Center for Astrophysics, California Institute of Technology, 1216 East California Boulevard, Mail code 278-17, Pasadena California 91125, USA}

\begin{abstract} 
Dark-matter-only simulations predict that dark matter halos have cusp-like inner density profiles, while observations of low-mass galaxies have found a range of inner slopes that are typically much shallower. It is still not well established whether this discrepancy can be explained by baryonic feedback or if it may require modified dark matter models. To better understand the diversity of dark matter profiles in dwarf galaxies, we undertook a survey of 26 low-mass galaxies ($\log M_*/\msol = 8.4-9.8$, $v_{\rm max} = 50-140$ km s$^{-1}$) within 30 Mpc using the Palomar Cosmic Web Imager, which is among the largest integral field spectroscopic surveys of its type. In this paper, we derive H$\alpha$ velocity fields for the full sample with a typical spatial resolution of $\sim$160 pc. We extract rotation curves and verify their robustness to several choices in the analysis. We present a method for improving the velocity precision obtained from image slicing spectrographs using narrowband H$\alpha$ images. For 11 galaxies, we compare the H$\alpha$ velocity fields to CO kinematics measured using CARMA, finding the maps to be in good agreement. The standard deviation of the difference is typically $\sim$7 km s$^{-1}$, comparable to the level of turbulence in the interstellar medium, showing that the two tracers have substantially the same bulk kinematics. In a companion paper, we will use the rotation curves produced here to construct mass models of the galaxies and determine their dark matter density profiles. 
\end{abstract}

\section{Introduction}\label{Intro}

The $\Lambda$ cold dark matter ($\Lambda$CDM) model has been a powerful tool for studying a wide range of phenomena. It has successfully reproduced many of the observable properties of the universe over a variety of scales. There are, however, still challenges facing the $\Lambda$CDM model, and in particular several small-scale discrepancies (see \citealt{Bullock} for a review). An important question is whether these apparent discrepancies arise from baryonic effects in galaxy evolution or whether they result from dark matter microphysics. 

$N$-body simulations were famously used to investigate the density profile of dark matter halos by \cite{NFW}, who found that a single universal density profile can describe dark matter halos over a wide range of masses. This Navarro-Frenk-White (NFW) profile, despite its usefulness and range of applications, is the source of some of the challenges to $\Lambda$CDM. In particular, the so-called ``cusp-core" problem stems from a discrepancy in the central region of galaxies between the observed dark matter density profile and the standard NFW profile found from $N$-body simulations. According to the NFW profile, the inner slope of the dark matter distribution should scale as $\rho \propto r^{-1}$. Instead of this `cuspy' distribution in the center, observations have found that dwarf galaxies can have shallower profiles where $\rho \propto r^{-\beta}$ with $\beta < 1$ (see \citealt{deBlokReview} for a review). Dark matter profiles that deviate significantly from NFW are generally referred to as ``cores". Low-mass galaxies are excellent candidates for studying this problem because they are relatively dark matter dominated. 

Attempts to explain the presence of cores have led to the emergence of three main potential solutions: baryonic feedback, modified dark matter models, and structural features of the galaxy. Baryonic feedback, in particular from supernovae, could potentially drive dark matter particles away from the center via the change in gravitational potential that occurs when gas is expelled \citep{NFWfeedback, Mashchenko, Governato, Pontzen2012, Pontzen2014, Chan}. Modified CDM models, such as allowing self-interactions, could also be responsible for moving the particles into a cored distribution \citep{Kaplinghat, SIDM1, Peter, Rocha, Ren}, though warm dark matter is unlikely to erase cusps \citep{Maccio}. Bars or baryonic clumps in the structure of a galaxy may contribute to the flattening of the density profile through dynamical effects \citep{Weinberg, Katz, Holley, Tonini, Rand}. There is still much debate about which processes ultimately control the inner dark matter distribution in galaxies, especially whether physics beyond collisionless CDM is needed. 

One step toward resolving this question is obtaining a more detailed observational picture. Rotation curves are one of the earliest ways that the existence of dark matter was inferred (case of M31: \citealt{Rubin}; case of NGC 300 and M33: Appendix A of of \citealt{Freeman}) and still represent a valuable tool for exploring its distribution. Gas kinematics are used to trace the rotational velocity of the galaxy at various radii, and by modeling the resulting curve and the contribution of the baryon distribution, one obtains the dark matter density profile. This has been the primary method used to obtain the observational measurements that led to the cusp-core problem, and various groups have used it to further investigate the discrepancy. 

After the first indications of the cusp-core problem \citep{Flores, Moore}, observers used a variety of observational techniques to measure the rotation curves of low-mass galaxies, which have attracted most of the attention, due to their dominant dark matter component. \ion{H}{1} data were used by \cite{deBlok97} and eventually led to The \ion{H}{1} Nearby Galaxy Survey (THINGS; \citealt{things}), as well as a second survey, LITTLE THINGS \citep{littlethings}, which focused on dwarf galaxies. These surveys found that cored density profiles are prevalent in their samples \citep{thingsresults, Oh2011, thingsresults2, Oh2015}. Studies based on long-slit optical spectroscopy of ionized gas kinematics \citep{deblok01, mcgaugh, swaters} found further evidence that rotation curves are better fit with noncuspy density profiles. Groups using integral field spectroscopy \citep{simon, simon05, kuzio06, kuzio08, Diskmass7, Adams, Lelli1, Lelli2, Korsaga1, Korsaga2} generally find shallower-than-NFW dark matter profiles as well, though they find a wider diversity of density profiles, indicating that the picture may be more complicated. 

It thus seems clear that low-mass galaxies often, but not always, have dark matter density profiles flatter than the NFW profile. However, the variation in these profiles is not well understood and will be key to understanding the physical mechanisms that govern the observed profiles \citep{Diversity-Oman}. With this in mind, we have undertaken a survey of dwarf galaxy kinematics to better constrain the origin and diversity of their dark matter profiles, with the goal of ultimately correlating these profiles with other galaxy properties. 

We have observed 26 galaxies using integral field spectroscopy covering the H$\alpha$ line. Eleven of these galaxies have also been studied as part of a parallel observational campaign using the Combined Array for Research in Millimeter-wave Astronomy (CARMA) to trace their CO kinematics \citep{Mai1}. The survey provides several advantages over earlier work. First, 2D kinematics improve on long-slit samples by avoiding the slit-misplacement issues and allow us to diagnose the level of noncircular motions that may lead to spurious inferences of cores. Second, the $\sim 2\arcsec$ resolution of the H$\alpha$ observations improves on \ion{H}{1} observations and much of the earlier H$\alpha$ integral field unit (IFU) work based on DensePak/SparsePak. Third, our sample of 26 galaxies is large enough to begin to characterize the distribution of dark matter profiles and examine correlations with the baryonic properties of the galaxies. It is among the largest samples of optical velocity fields of dwarf galaxies (see also \citealt{ghasp}). Fourth, we are able to compare the ionized and molecular gas kinematics for the largest sample of dwarf galaxies to date and verify that results on the dark matter distribution do not depend on the dynamical tracer.

For our survey, we chose to use the Palomar Cosmic Web Imager (PCWI), an IFU that has a spectral resolving power of $R \sim5000$ and an angular resolution of $\sim 2\arcsec$ \citep{CWI}. The improved angular sampling also allows us to select galaxies that are farther away while maintaining adequate spatial resolution, which enables our large sample. A large sample gives us the ability to measure scatter in the inner slope of the dark matter density profiles, and we will additionally be able to look for correlations of the slope with other galaxy properties. Such correlations can be compared to predictions made by baryonic feedback models \citep{Chan, FIRE} and self-interacting dark matter (SIDM) models \citep{Ren} and could help better constrain the origin of cores in low-mass galaxies. 


This paper is the first of two on the results of the H$\alpha$ survey. It presents our galaxy sample selection, observations and data reduction, comparison to molecular gas velocity fields, photometry, and the creation of rotation curves. Paper II will present our mass models, the inner dark matter density slopes, and our interpretation of these dark matter profiles based on comparisons to other samples and theoretical work. 

The paper is organized as follows: Section 2 covers our sample selection, details about our H$\alpha$ observations with PCWI, and comparison to the CO velocity fields. Section 3 describes our photometric data and measurements. Section 4 describes the process of deriving the rotation curves. In Section 5 we compare our rotation curves to those in the literature. Finally, Section 6 summarizes the results of this paper.

\section{Observations with PCWI}\label{CWI}

\subsection{Galaxy Sample Selection}\label{sample}
Explicit details and criteria used to select our targets can be found in \cite{Mai1} but are summarized here. We first selected low-mass galaxies by considering those with \ion{H}{1} line widths $W_{20} \lesssim 200$ km s$^{-1}$ and absolute magnitude $M_B \gtrsim -18$. Second, the galaxies were required to have an inclination between $30^{\circ}$ and $70^{\circ}$ and a diameter larger than $1\arcmin$. Third, we considered those that were late type and generally symmetric and regular in shape, with no obvious central bar or distortion. We also limited our search to galaxies within 30 Mpc. Given our expected resolution of about $2\arcsec$, the innermost radius we could hope to observe is around $1\arcsec$, which at the maximum distance of 30 Mpc corresponds to 145 pc, and to 78 pc at our sample's median distance of 16 Mpc. 

Our final sample consists of 26 galaxies with typical luminosities of $10^9$ L$_\sun$ in the $r$ band and rotation velocities of around 50 - 140 km s$^{-1}$. Two galaxies, UGC 3371 and UGC 11891, are classified as low surface brightness (LSB) galaxies. Twenty-two of our galaxies overlap with the 26 targeted for CO observations by \cite{Mai1}, though of their 26 only 14 had sufficiently bright and extended CO to produce viable rotation curves, and of those we overlap with 11 galaxies. We compare the H$\alpha$ and CO velocity fields in Section \ref{CO_comp} and will examine the inner dark matter slopes in Paper II. 

The names and properties of our 26 H$\alpha$ galaxies can be found in Table \ref{proptable}. To determine distances, we found that 13 of our galaxies have Tully-Fisher distances from \cite{Tully}. Of the remaining galaxies, we used Tully-Fisher distances from \cite{Gavazzi} for NGC 4376 and NGC 4396 and from \cite{3371} for UGC 8516. For the rest, we used the Hubble flow distance with $H_0 = 73.0$ and corrected for large-scale flows toward the Virgo Cluster, the Shapley Supercluster, and the Great Attractor, as listed in the NASA/IPAC Extragalactic Database (NED).

\begin{deluxetable*}{lccccccc} 
\tablecolumns{8}
\tablewidth{0pt}
\tablecaption{Sample Galaxy Properties}
\tablehead{\colhead{Name} & \colhead{R.A.} & \colhead{Decl.} &  \colhead{Distance} & \colhead{$V_{\rm sys}$} & \colhead{$\log(L_r)$}& \colhead{$\log(L_{Ch2})$} \\ \colhead{} & \colhead{} & \colhead{} & \colhead{(Mpc)} & \colhead{(km s$^{-1}$)} & \colhead{} & \colhead{}} 
\startdata
NGC 746 & 29.46294 & 44.91834 & 12$^b$ & 702 & 8.97 & - \\ 
NGC 853$^*$ & 32.92178 & -9.30524 & 20$^b$ & 1518 & 9.48 & - \\ 
NGC 949$^*$ & 37.70259 & 37.13668 & 10$^a$ & 605 & 9.28 & - \\ 
NGC 959 & 38.09967 & 35.49463 & 10$^a$ & 593 & 9.01 & 9.49 \\ 
NGC 1012$^*$ & 39.81184 & 30.15114 & 14$^b$ & 969 & 9.43 & - \\ 
NGC 1035$^*$ & 39.87136 & -8.13283 & 16$^a$ & 1240 & 9.49 & 10.26 \\ 
NGC 2644 & 130.38355 & 4.98104 & 30$^b$ & 1932 & 9.75 & 10.33 \\ 
NGC 2976 & 146.81385 & 67.91668 & 4$^a$ & 1 & 8.35 & 9.58 \\ 
NGC 3622 & 170.05169 & 67.24187 & 23$^a$ & 1317 & 9.46 & 9.9 \\ 
NGC 4376 & 186.32526 & 5.74134 & 24$^c$ & 1132 & 9.39 & 9.82 \\ 
NGC 4396 & 186.49512 & 15.67166 & 16$^c$ & -120 & 9.31 & 9.89 \\ 
NGC 4451$^*$ & 187.16880 & 9.25915 & 26$^a$ & 849 & 9.74 & 10.31 \\ 
NGC 4632$^*$ & 190.63331 & -0.08242 & 14$^a$ & 1702 & 9.48 & 10.15 \\ 
NGC 5303$^*$ & 206.93757 & 38.30463 & 28$^b$ & 1421 & 9.66 & 10.34 \\ 
NGC 5692$^*$ & 219.57549 & 3.41031 & 27$^b$ & 1585 & 9.57 & - \\ 
NGC 5949$^*$ & 232.00273 & 64.76313 & 13$^a$ & 434 & 9.35 & 9.94 \\ 
NGC 6106$^*$ & 244.69664 & 7.41082 & 24$^a$ & 1443 & 9.81 & 10.38 \\ 
NGC 6207$^*$ & 250.76550 & 36.83213 & 16$^a$ & 843 & 9.67 & 10.23 \\ 
NGC 6503 & 267.36004 & 70.14434 & 6$^a$ & 29 & 9.47 & 10.1 \\ 
NGC 7320 & 339.01423 & 33.94826 & 14$^b$ & 771 & 9.14 & 9.63 \\ 
UGC 01104 & 23.17725 & 18.31687 & 10$^b$ & 680 & 8.42 & 8.76 \\ 
UGC 3371 & 89.15165 & 75.31691 & 15$^b$ & 814 & 8.88 & - \\ 
UGC 4169 & 120.63655 & 61.38821 & 30$^a$ & 1603 & 9.62 & 10.04 \\ 
UGC 8516 & 202.96847 & 20.00127 & 22$^d$ & 1021 & 9.27 & 9.77 \\ 
UGC 11891 & 330.89073 & 43.74893 & 8$^a$ & 463 & 9.11 & - \\ 
UGC 12009 & 335.66767 & 37.97735 & 20$^b$ & 1221 & 9.26 & - 
\enddata
\tablecomments{Galaxies marked with $*$ are those that overlap the viable CO sample from \cite{Mai1}. Distances are marked to indicate sources as follows (see Section \ref{sample}): $a$ - \cite{Tully}, $b$ - Hubble flow distance, $c$ - \cite{Gavazzi}, $d$ - \cite{3371}. Values of $V_{\rm sys}$ are taken from our analysis (see Section \ref{DiskFit}) with typical random uncertainties of 1-2 km s$^{-1}$. The remaining parameters were derived from photometry, as described in Section \ref{photometry}. For galaxies with a tabulated IRAC Channel 2 luminosity, we will preferentially use infrared photometry in our analysis. \label{proptable}}
\end{deluxetable*}

\subsection{PCWI Observations}
Since we ultimately would like to probe the inner dark matter distributions of these galaxies, our observations require an instrument with a sufficiently high spatial and spectral resolution to measure gas kinematics near their centers. We chose to use PCWI, an integral field spectrograph at Palomar Observatory. It is mounted at the Cassegrain focus of the Hale 5m telescope and has a resolving power of $R \sim5000$  \citep{CWI}. An image slicer divides the approximately $60\arcsec$x$40\arcsec$ field of view into 24 slices, each approximately $40\arcsec$x$2\farcs6$. The spatial resolution is limited by the seeing on the $40\arcsec$ axis (along each slice) and by the $2\farcs6$ slice width along the perpendicular axis, although this limit can be improved by dithering between exposures, as described below. 

The 26 galaxies in our sample were observed during 17 nights from 2013 to 2015. To locate each galaxy for the exposures, a nearby bright star was centered in the field of view, and then the telescope was offset to the position of the galaxy. For the pointing covering the center of each galaxy, where the highest possible spatial resolution is necessary, we obtained consecutive exposures with the telescope dithered by $1\farcs3$, half the slice width, in the direction perpendicular to the slices. By combining these interlaced exposures as described in the following section, we aim to recover some of the spatial information that would otherwise be lost due to the slices undersampling the seeing. The mean seeing for each observation date ranges from $1\farcs3$ to $2\farcs7$, with an overall sample mean of $1\farcs8$. 

The majority of our galaxies were observed using the nod and shuffle technique. This method, developed by \cite{NandS}, allows for high-precision sky subtraction by regularly taking sky exposures close in time and position to the target, nodding the telescope without reading out the detector.  A mask is used to block the top and bottom portions of the detector so that only the middle third is illuminated. For each galaxy, a blank sky position is selected and a nearby star is chosen for guiding in both the target and sky regions. The telescope is nodded between the target and sky fields periodically, with a coordinated shuffle of charge on the CCD so that the illuminated portion of the detector (the middle third) contains the appropriate target or sky exposure. The nod and shuffle technique greatly reduces read noise while allowing us to sample the temporal variations in the sky background. The process ultimately produces two spectra, one of the target with sky and one of only the sky, which can easily be subtracted to obtain the target spectrum. 

When using this technique, we nodded between sky and target every 2-3 minutes for 10-20 minutes of total exposure time. We also obtained calibration exposures using the internal quartz and arc lamps regularly throughout the night. Since nod and shuffle requires the use of a nearby guide star, for targets where no such star was present or conditions prevented us from seeing the star, we could not use the technique. This was the case for five of our galaxies, and for these we instead took separate sky exposures to use for subtraction. 

In most cases, the galaxy was larger than the PCWI field of view, so exposures had to be taken at several positions to cover it. These groups of exposures, referred to as tiles, are later stitched together to create a single image (see Section \ref{DR}).

Details about the observations of each galaxy can be found in Table \ref{obstable}. 

\begin{deluxetable*}{lcccc} 
\tablecolumns{5}
\tablewidth{0pt}
\tablecaption{Observation Details}
\tablehead{\colhead{Name} & \colhead{Obs. Date(s)} & \colhead{No. of Tiles} & \colhead{Mean Seeing ($\arcsec$)} & \colhead{Nod and Shuffle?}} 
\startdata
NGC 746 & 2013 Nov 27 & 4 & 1.85 & Yes \\
NGC 853 & 2014 Aug 20 & 2 & 1.25 & Yes\\
NGC 949 & 2013 Sep 30 & 4 & 1.75 & Yes\\
 & 2013 Nov 27 & & & \\
NGC 959 & 2013 Nov 28 & 4 & 1.52 & Yes\\
NGC 1012 & 2014 Aug 20 & 4 & 1.41 &Yes\\
NGC 1035 & 2013 Nov 29 & 5 & 2.11 &No\\
 & 2013 Nov 30& & & \\ 
NGC 2644 & 2015 Feb 26 & 3 & 2.20&Yes \\
NGC 2976 & 2015 Mar 15 & 9 & 1.82& Yes\\
 &2015 Mar 16& & & \\
NGC 3622 & 2014 Nov 18 & 2 & 2.05 & Yes\\
NGC 4376 & 2014 May 29 & 4 & 1.61 & No \\
NGC 4396 & 2015 Feb 26& 6 & 1.80 & Yes \\
 & 2015 March 15 & & & \\
NGC 4451 & 2015 Mar 16 & 2 & 1.94 & Yes\\
NGC 4632 & 2015 Mar 16 & 4 & 1.42 & Yes\\
NGC 5303 & 2014 Mar 30 & 4 & 1.98 & No\\
NGC 5692 & 2015 Mar 16 & 3 & 1.53 & Yes\\
NGC 5949 & 2014 Aug 20 & 4 & 1.86 & Yes\\
NGC 6106 & 2014 May 29 & 7 & 1.56 & Yes\\
NGC 6207 & 2014 Aug 20 & 8 & 1.75 & Yes\\
 & 2014 Aug 21& & & \\
NGC 6503 & 2014 Aug 22 & 10 & 1.71 & Yes\\
NGC 7320 & 2013 Sep 30 & 4 & 1.26 & Yes \\
UGC 01104 & 2014 Nov 17 & 2 & 2.69 & Yes\\
UGC 3371 & 2013 Nov 27  & 13 & 2.32 & No\\
 & 2014 Aug 21& & & \\
  & 2014 Aug 22& & & \\
   & 2014 Nov 17& & & \\
UGC 4169 & 2014 Nov 17 & 4& 2.49 & Yes\\
 & 2014 Nov 18& & & \\
UGC 8516 & 2015 Mar 15 & 3 & 1.38 & Yes \\
UGC 11891 & 2013 Oct 1& 16  & 1.95 & No \\
 & 2014 Nov 17 & & & \\
UGC 12009 & 2013 Sep 30 & 2 & 1.34 & Yes
\enddata
\tablecomments{The tile number includes both the center and the dithered image of the center. \label{obstable}}
\end{deluxetable*}

\subsection{Data Reduction}\label{DR}
Individual exposures were reduced by using the PCWI data reduction pipeline \citep{Coldplay}. The pipeline performs bias subtraction, sky subtraction for exposures using nod and shuffle, flat-fielding, derivation of the wavelength solution, and rectification. This process produces a data cube for each exposure. 

The data cubes are then further processed through a custom set of IDL routines designed to produce maps of the H$\alpha$ intensity, velocity, and velocity uncertainty. A zero-point correction to the wavelength solution is determined for each image using night-sky emission lines. This correction accounts for instrumental flexure between the science observations and the calibrations. For observations without the nod and shuffle technique, the intensity of skylines and flexure changes over time, which precludes a straightforward subtraction of the sky exposure. Instead, we shift the sky spectra slightly in wavelength and rescale the intensity, adjusting until the residuals are minimized. During the reduction, the PCWI slices are binned along their length to create spaxels that are $1\farcs3$x$2\farcs6$. The H$\alpha$ line in each spaxel is then fit with a Gaussian profile, which  we found to adequately describe the emission. 

Our procedure for acquiring the galaxies during observation was not accurate to arcsecond precision, so the coordinate map produced by the PCWI pipeline must be adjusted. To do this, we compare the distribution of H$\alpha$ in each data cube to a narrowband image. More details about the imaging observations are given in Section \ref{photometry}. The H$\alpha$ intensity map produced from the PCWI data cube was cross-correlated in 2D with the narrowband H$\alpha$ image to derive the absolute coordinates.  

In cases where it took more than one pointing to cover the galaxy, the 2D flux and velocity fields are stitched together to form a single field. The process used is similar to the DRIZZLE method, introduced by \cite{drizzle}. The separate exposures are drizzled onto a grid of $1\farcs32$ pixels, which is approximately half of a slice. The dithering method used in the central pointing similarly steps by half of a slice, which allows for improved resolution in the center. 

\begin{figure*}
\centering
\includegraphics[width=1.0\textwidth]{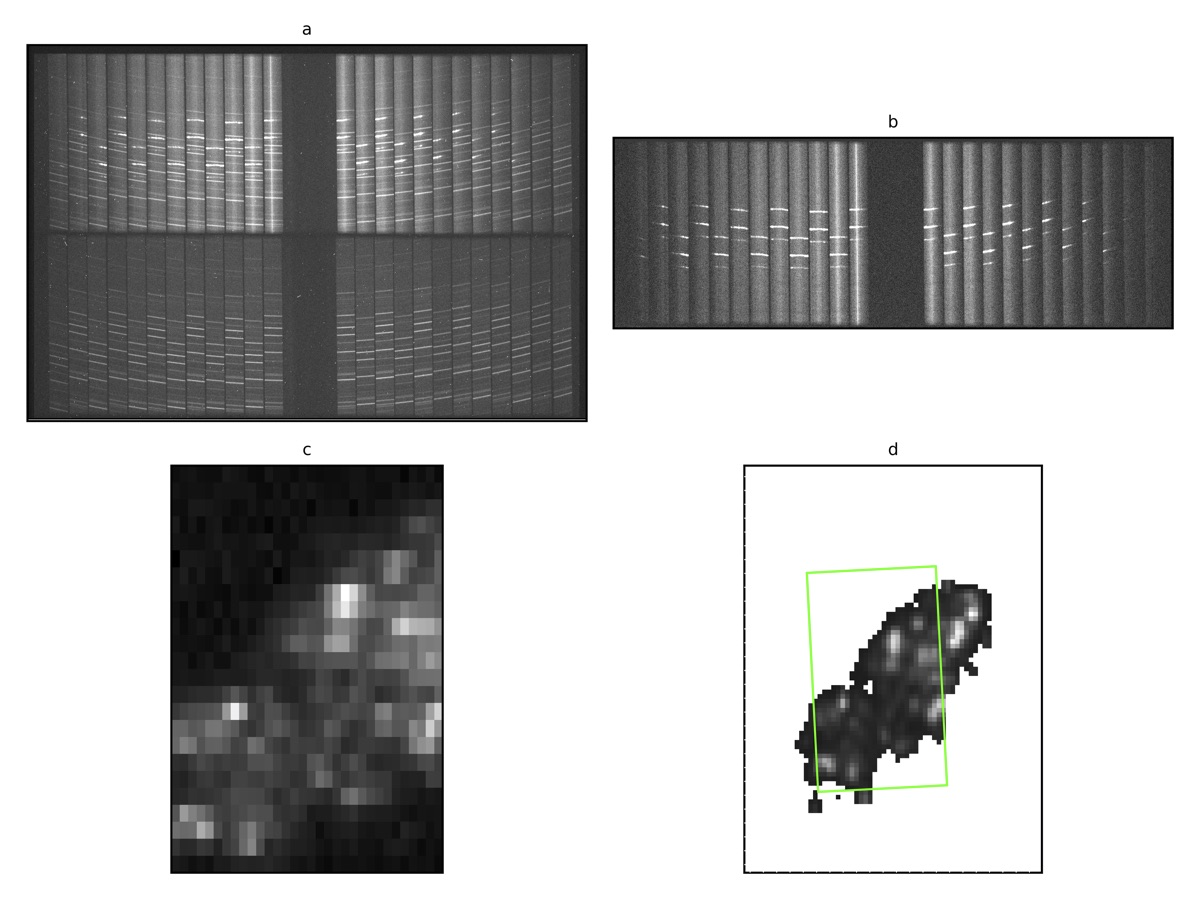}
\caption{Illustration of several stages of data reduction. (a) Raw PCWI image, object spectrum on top and sky spectrum below. (b) Sky-subtracted, flat-field-corrected, object spectrum. (c) H$\alpha$ intensity field produced from the reduced data cube. (d) Total H$\alpha$ field of the galaxy (NGC 5949) after combining four exposures. The region corresponding to panel (c) is highlighted by the green box, which represents the $60\arcsec$x$40\arcsec$ PCWI field of view . Note that the pixel scale in panel (c) is $2\farcs6$ in the vertical axis and $1\farcs3$ in the horizontal direction.}
\label{fig:cwifig}
\end{figure*}

\begin{figure*}
\centering
\includegraphics[width=1.0\textwidth]{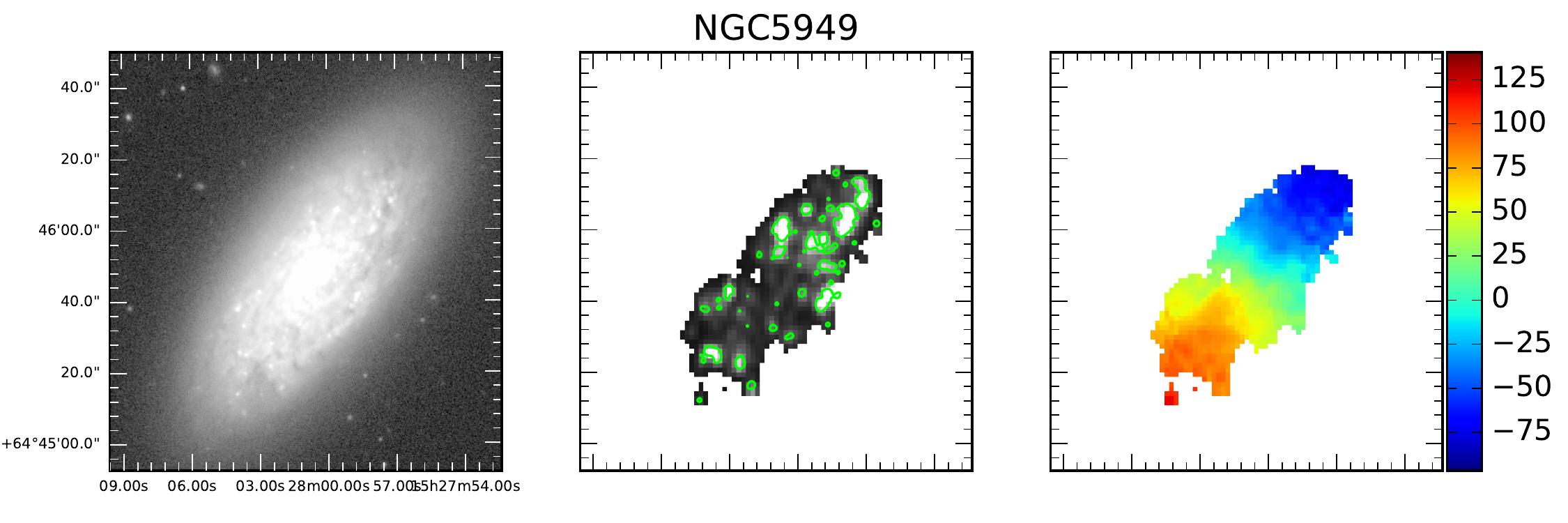}
\caption{From left to right: the $r$-band image, total H$\alpha$ flux from PCWI (with contours from photometry in green), and velocity field derived from the PCWI data with $V_{\rm sys}$ removed. Corresponding images for each of the 26 galaxies can be found in the bottom panels of the Appendix figures. }
\label{fig:maps}
\end{figure*}

Various steps in the data reduction process for a sample galaxy (NGC 5949) are displayed in Figure \ref{fig:cwifig}, while an $r$-band image, the final H$\alpha$ flux map, and the velocity field for the same galaxy can be found in Figure \ref{fig:maps}. 

Due to the angular resolution and sensitivity of PCWI, we generally detect H$\alpha$ uniformly throughout the disk, ensuring highly detailed velocity fields. The observations sample almost the full range of azimuth at most locations, which is crucial for measuring noncircular motions. As we average over annuli, these patches will not have a significant effect on the derived rotation curves. The H$\alpha$ flux maps and velocity fields for all galaxies can be found in the bottom middle and bottom right panels of the Appendix figures.

\section{Imaging Data and Analysis}\label{photometry}
We require photometric observations to determine many of the galaxy properties given in Table \ref{proptable} and to assist in the reduction of the PCWI data. We additionally will use photometry to estimate the stellar contribution to the rotation curve, which will be used in mass modeling in Paper II. We therefore took additional observations of each galaxy from various sources and modeled the stellar light distribution. 

\subsection{Photometric Observations}\label{phot}
For each galaxy we require an $r$-band image and a narrowband image around the H$\alpha$ line.  Continuum emission is removed from the narrowband image to produce the maps of H$\alpha$ that are used to register the PCWI exposures to absolute coordinates. We observed 21 of our galaxies using SPICAM on the 3.5m telescope at Apache Point Observatory (APO) over three nights: 2013 October 5, 2014 March 5, and 2014 November 27. For four of the remaining galaxies (NGC 2976, NGC 6503, UGC 3371, UGC 11891), the $r$-band and H$\alpha$ data were taken from the Local Volume Legacy Project. These observations were obtained with the Steward Observatory Bok 2.3m telescope on Kitt Peak, as described in \cite{LVL}. For the remaining galaxy (NGC 2644), we used images taken from the 1.0m Jacobus Kapteyn Telescope on La Palma as part of the survey done by \cite{LaPalma}. We masked other galaxies and stars in the field of view by using Source Extractor \citep{source} to locate them and place elliptical masks. The $r$-band images can be seen in the bottom left panels of the Appendix figures, while contours from the narrowband images are overlaid in the bottom middle panels. 

For 18 of the galaxies, we have additional images from Channel 2 (4.5 $\mu$m) of \textit{Spitzer's} Infrared Array Camera (IRAC; \citealt{IRAC}), obtained from the \textit{Spitzer} Heritage Archive and the \textit{Spitzer} Survey of Stellar Structure in Galaxies \citep{Spitzer}. For the analysis described below, if a galaxy had both infrared and $r$-band observations, we preferentially used the infrared images to define the parameters, as they more closely trace the stellar mass distribution \citep{Korsaga1, Korsaga2}. 

\subsection{Galaxy Centers and Ellipticities}
We use two independent procedures to estimate the center, ellipticity, and PA of our galaxies. Comparison of the results both helps ensure accuracy by verifying agreement and better addresses special cases where one method may be better suited. We first use a program called {\tt find$\_$galaxy} that is included in the multi-Gaussian expansion (MGE) package presented by \cite{MGE}. We then use the {\tt ELLIPSE} task in IRAF to compare and verify the MGE values. If the galaxy has an obvious nucleus, we chose the coordinates of the nucleus to define the center rather than using a fitting procedure. 

The {\tt find$\_$galaxy} method finds the center, ellipticity, and PA by computing a best-fit ellipse using the moments of the pixels above a given surface brightness threshold (the results were not very sensitive to the threshold). The IRAF {\tt ELLIPSE} procedure instead fits ellipses to a sequence of isophotes. This allows us to test the variation of the ellipticity with radius and is much less susceptible to irregularities in the central region, so it is more appropriate for galaxies with bright or chaotic centers. In the cases of NGC 746, NGC 853, and NGC 1012, we chose to use {\tt ELLIPSE} values over {\tt find$\_$galaxy} values because of their irregular central regions. This is generally a consequence of using $r$-band data, which is not as smooth as the 4.5$\mu$m data, rather than a reflection of abnormal behavior at the center. For all other cases, the {\tt ELLIPSE} and {\tt find$\_$galaxy} values agreed to $\sim0.02$ in ellipticity and $1^{\circ}$ in PA. These estimates are needed to extract rotation curves from the PCWI data (see Section \ref{DiskFit}). In Paper II, we will analyze the surface brightness profiles obtained from these images in order to estimate the contribution of stellar mass to the rotation curve.

\section{Rotation Curves Using DiskFit}\label{DiskFit}

To create the rotation curves, we used DiskFit to model the velocity fields produced from the PCWI data. DiskFit is a publicly available program written by \cite{SandS} to fit models to images or velocity fields of disk galaxies, which is done by minimizing the $\chi^2$ estimate of the differences between a projected model and the given input data. It outputs estimates of the radial and tangential velocities at a sequence of radii provided by the user. While DiskFit is capable of fitting more complicated nonaxisymmetric models, we chose to initially fit our velocity fields using the purely axisymmetric radial flow model, given by \\ 
\begin{equation}
V_{model} = V_{sys} + \sin i (\bar{V_t} \cos \theta + \bar{V_r} \sin \theta)\\
\end{equation}
where $V_{\rm sys}$ is the systemic velocity, $i$ is the inclination, $\theta$ is the angle with respect to the major axis, and $V_t$ and $V_r$ are the tangential and radial velocities, respectively. In this model, the output tangential and radial velocities are constant on ellipses defined by a common center, ellipticity, and PA. Initial estimates of these values from photometry are provided to the program and are either kept fixed throughout or included in the fitting procedure as free parameters. 

DiskFit also has the capability to fit a velocity field with a bisymmetric distortion that can be induced by a bar. In this case, the model is given by 
\begin{multline}\label{bisym-eq}
V_{model} = V_{sys} + \sin i [\bar{V_t} \cos \theta - V_{2,t} \cos (2\theta_b) \cos \theta \\
- V_{2,r} \sin (2\theta_b) \sin \theta]  \\
\end{multline}
where a subscript of 2 indicates the second harmonic and $\theta_b$ is the position angle of the distortion. We also fit our data with this model, and we will explore evidence of bisymmetric distortions in Section \ref{bar}.

\begin{figure*}
\centering
\includegraphics[width=1.0\textwidth]{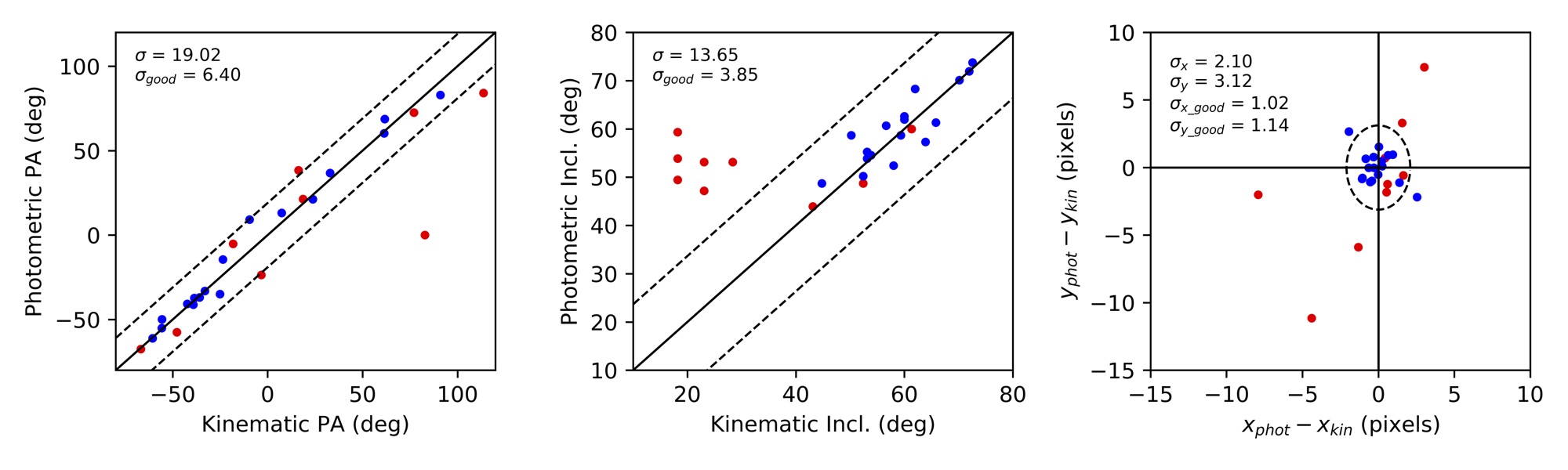}
\caption{Comparison of photometric and kinematic fit values for PA, inclination (derived from ellipticity), and center for our sample. Red points represent galaxies that we restrict to photometric values owing to poor kinematic fit results. Solid lines are the one-to-one line, and the dashed lines show one standard deviation in either direction. The standard deviation of the total sample $\sigma$ is given, as well as that of only the blue (good) points $\sigma_{\rm good}$. }
\label{fig:PvsK}
\end{figure*}

\begin{deluxetable*}{lcccccccc} 
\tablecolumns{9}
\tablewidth{0pt}
\tablecaption{Parameters: Photometric vs. Kinematic (DiskFit)}
\tablehead{\colhead{Name} & \colhead{$x_{\rm phot} - x_{\rm kin}$}  &\colhead{$y_{\rm phot} - y_{\rm kin}$} & \colhead{Phot. PA} & \colhead{Kin. PA} & \colhead{Phot. $e$} & \colhead{Kin. $e$}  &  \colhead{Phot. Incl} &  \colhead{Kin. Incl.}   \\ \colhead{} & \colhead{(arcsec)} & \colhead{(arcsec)} & \colhead{(deg)} & \colhead{(deg)} & \colhead{} & \colhead{} & \colhead{(deg)} & \colhead{(deg)} }
\startdata
NGC 746$^*$& -1.3 & -5.9 & 84.1 & 113.7 & 0.35 & 0.05 & 49.5 & 18.2 \\ 
NGC 853$^*$& 0.4 & 0.7 & 72.5 & 77.0 & 0.32 & 0.08 & 47.2 & 23.5 \\ 
NGC 949& -0.3 & -0.0 & -37.2 & -38.7 & 0.48 & 0.36 & 58.7 & 49.9 \\ 
NGC 959& -1.1 & -0.8 & 68.7 & 61.7 & 0.39 & 0.47 & 52.4 & 58.1 \\ 
NGC 1012$^*$& 1.6 & 3.3 & 21.4 & 18.6 & 0.49 & 0.05 & 59.3 & 18.2 \\ 
NGC 1035& 0.2 & 0.5 & -33.1 & -33.0 & 0.69 & 0.69 & 71.9 & 72.1 \\ 
NGC 2644& -1.9 & 2.7 & 13.1 & 7.4 & 0.54 & 0.50 & 62.6 & 59.8 \\ 
NGC 2976& -0.5 & -1.1 & -34.9 & -25.1 & 0.52 & 0.59 & 61.3 & 65.8 \\ 
NGC 3622& 0.0 & 1.5 & 9.2 & -9.5 & 0.42 & 0.41 & 54.5 & 53.8 \\ 
NGC 4376$^*$& 1.6 & -0.6 & -23.6 & -3.3 & 0.40 & 0.12 & 53.1 & 28.1 \\ 
NGC 4396& 1.4 & -1.1 & -55.1 & -55.7 & 0.72 & 0.70 & 73.7 & 72.6 \\ 
NGC 4451& -0.3 & 0.8 & -14.4 & -23.6 & 0.34 & 0.29 & 48.7 & 45.1 \\ 
NGC 4632& 0.2 & 0.1 & 60.2 & 61.4 & 0.63 & 0.53 & 68.3 & 62.1 \\ 
NGC 5303& 0.6 & 0.9 & 82.9 & 90.9 & 0.41 & 0.40 & 53.8 & 53.0 \\ 
NGC 5692& -0.8 & 0.6 & 36.8 & 32.8 & 0.36 & 0.39 & 50.2 & 52.4 \\ 
NGC 5949& -0.4 & -1.0 & -36.9 & -35.8 & 0.53 & 0.50 & 62.0 & 59.9 \\ 
NGC 6106& -1.1 & -0.8 & -40.7 & -42.4 & 0.43 & 0.40 & 55.2 & 53.4 \\ 
NGC 6207& -0.6 & -0.0 & 21.2 & 23.8 & 0.51 & 0.45 & 60.7 & 56.7 \\ 
NGC 6503& -0.0 & -0.5 & -61.1 & -60.6 & 0.66 & 0.66 & 70.1 & 70.4 \\ 
NGC 7320& 1.0 & 1.0 & -49.9 & -55.7 & 0.48 & 0.49 & 58.7 & 59.2 \\ 
UGC 01104$^{**}$& -7.9 & -2.0 & 0.1 & 82.8 & 0.41 & 0.05 & 53.8 & 18.2 \\ 
UGC 3371$^*$& 3.0 & 7.4 & -57.5 & -47.8 & 0.40 & 0.08 & 53.1 & 22.7 \\ 
UGC 4169& 2.6 & -2.2 & -41.2 & -39.1 & 0.46 & 0.56 & 57.3 & 64.0 \\ 
UGC 8516$^*$& 0.5 & -1.8 & 38.4 & 16.2 & 0.34 & 0.39 & 48.7 & 52.1 \\ 
UGC 11891$^*$& -4.4 & -11.1 & -67.4 & -66.8 & 0.28 & 0.27 & 43.9 & 42.8 \\ 
UGC 12009$^*$& 0.6 & -1.2 & -5.2 & -18.2 & 0.50 & 0.52 & 60.0 & 61.1 
\enddata
\tablecomments{Kinematic values are from DiskFit with all parameters allowed to vary. Galaxies marked with an asterisk were then restricted to the photometric values for the remaining analysis owing to poor kinematic fits (see Figure \ref{fig:PvsK}), while those marked with double asterisks were restricted to the photometric values except for the PA (see Section \ref{special}). A negative in the the $x$-coordinate column indicates east, and a negative in the $y$-coordinate column indicates south. For inclination not explicitly fit with DiskFit, these values are derived using $\arccos(1-e)$. \label{PKtable}}
\end{deluxetable*}

\subsection{Fitting Assumptions}\label{ISM}
DiskFit requires, as input, the velocity field and its associated error. The formal velocity measurement uncertainties associated with fitting a Gaussian to an observed emission line are very small.  Given small-scale interstellar medium (ISM) motions that may vary randomly across a galaxy, they are also not representative of the true uncertainty in the rotation velocity as a function of radius. To account for random motions within the ISM, a value for $\Delta_{\rm ISM}$ is added to the velocity errors in quadrature when DiskFit is run. We chose $\Delta_{\rm ISM} $ = 6 km s$^{-1}$, as this was the value that produced a median reduced $\chi^2$ of 1 for the sample. This value is in agreement with that adopted by \cite{SandS} and \cite{Kuzio}. 

DiskFit fits velocities at specified radii. We chose to start at an inner radius of $r = 1$ pixel and then increase using a bin size of 2 pixels until we covered the entire H$\alpha$ field. Since 1 pixel is $1\farcs32$, $r=1$ approximately marks the smallest rescaled radius at which we can obtain measurements, given our typical seeing of $1\farcs8$. The cutoff point for the outermost radius varies by galaxy and was determined to be where the number of points contained in the annulus fell below 4.

\subsection{Disk Geometry}
As mentioned above, DiskFit also requires initial estimates of the shape, position, and systemic velocity of the galaxy, and these are either kept fixed or allowed to vary in the fitting procedure. For all parameters except $V_{\rm sys}$, we used values that were derived from our photometry, as described in Section \ref{photometry}. For $V_{\rm sys}$, our initial estimate was the value given by SIMBAD and was allowed to vary during fitting. Initially we allowed the disk geometry (center, ellipticity, and PA) to vary in the kinematic fit for all of the galaxies in our sample. For most galaxies this produces estimates very close to the photometric values; however, for about a third of our sample DiskFit ultimately fit values that were significantly different. Table \ref{PKtable} shows a comparison of the photometric and kinematic estimates of the ellipticity, PA, and center from DiskFit for each galaxy, while comparison plots for each parameter are provided in Figure \ref{fig:PvsK}. The most common issue was that the kinematic inclination was far too low. The corresponding rotation curves reflected these discrepancies, displaying highly nonmonotonic behavior and large radial motions or dramatic shifts with increased errors. Because of this, we decided to fix all three geometric parameters to their photometric values for galaxies exhibiting this behavior. For comparison, we also ran DiskFit on the remaining galaxies with fixed geometry parameters. 

\begin{figure}
\centering
\includegraphics[width=\columnwidth]{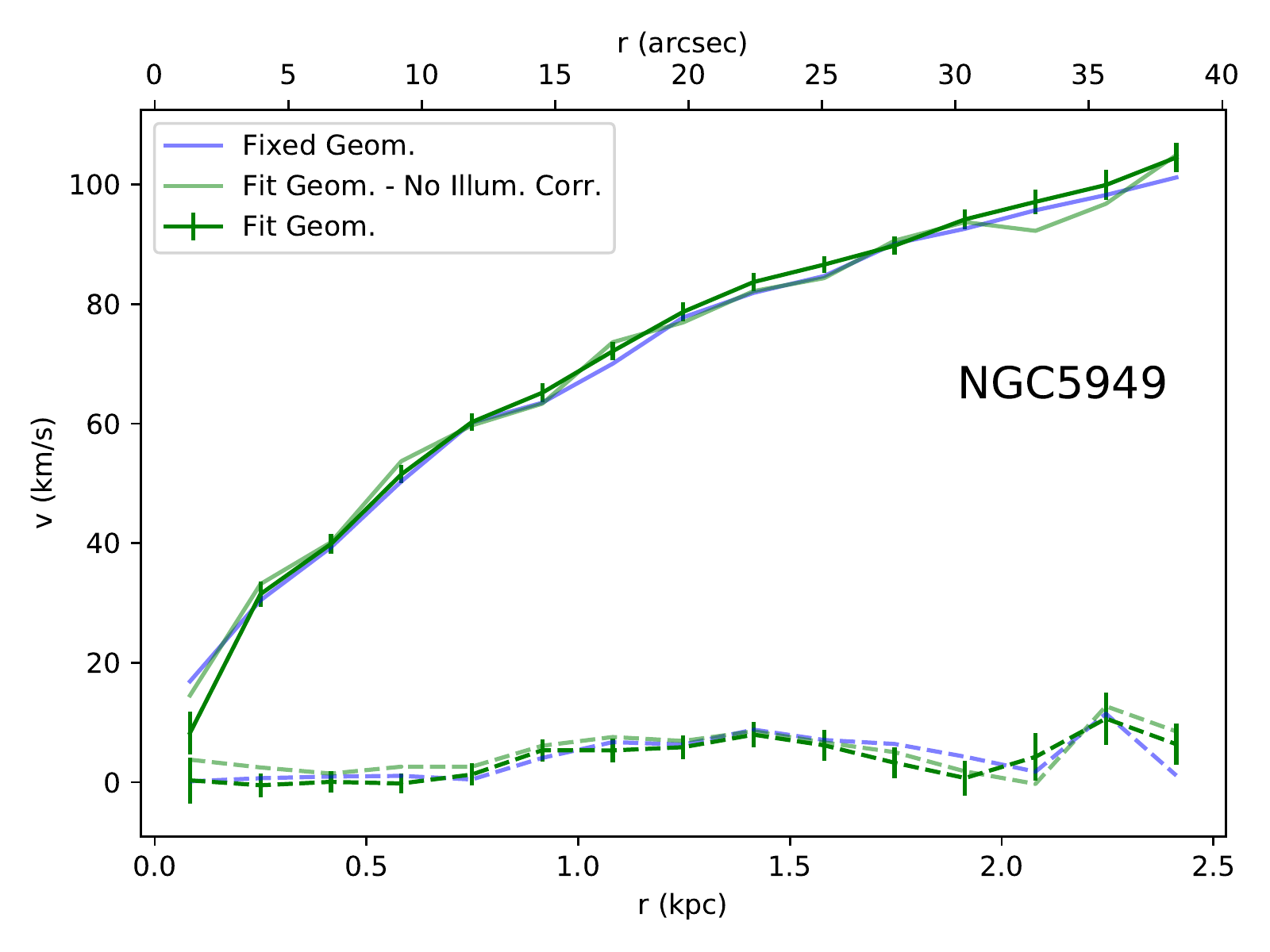}
\caption{Tangential (solid) and radial (dashed) rotation curves for NGC 5949. The curves generated with all parameters fixed to photometric values are shown in blue, and those for which all parameters were allowed to vary are shown in green. As described in Section \ref{velc}, we correct our measured velocities for H$\alpha$ slit illumination location. The rotation curves made with the uncorrected velocity fields (No Illum. Corr.) are shown behind the corrected ones with reduced opacity. Error bars are only shown on the curve that we will use for our analysis. Rotation curves for all 26 galaxies can be found in the top panels of the Appendix figures. \label{rotcurves}}
\end{figure}

Figure \ref{rotcurves} shows the tangential and radial rotation curves derived with all parameters fixed (Fixed Geom.) for NGC 5949 in blue, as well as those with all parameters allowed to vary (Fit Geom.) in green. Note that for the galaxies that converged in both cases, there is close agreement between the two methods of fitting. This can be seen in the top panels of the Appendix figures, which show the rotation curves for all galaxies in our sample. 

\subsection{Special Cases}\label{special}
Two galaxies, UGC 01104 and UGC 3371, required larger bin sizes of 5 pixels, due to their sparse H$\alpha$ distributions. UGC 3371 additionally displayed unphysical behavior from its innermost point, due to sparse H$\alpha$ emission, leading us to omit the first bin from the rotation curve. For our two LSB galaxies, UGC 3371 and UGC 11891, there were not adequate data to robustly determine the radial components $\bar{V_r}$. We therefore did not fit for radial velocities for these two galaxies. A similar problem occurred in just the outer part of NGC 6207, where the H$\alpha$ field becomes sparse. In this case, we fit for both tangential and radial velocities up to a radius of 4.2 kpc and set $\bar{V_r} = 0$ for the remainder of the galaxy. Finally, our photometric PA for UGC 01104 did not agree with the kinematic PA found with DiskFit. Upon inspection, it became clear that the photometric PA is not correct. We therefore chose to allow DiskFit to fit the PA for this galaxy, initializing it with the kinematic value. 

Several galaxies had a few isolated extreme outliers from the DiskFit model. We created normalized residual fields by subtracting the DiskFit model from the velocity field and dividing by the error. Bad pixels were defined as those with a normalized residual larger than 6 and were removed from the velocity field before refitting. Four galaxies required minor masking (an average of 2 pixels): NGC 1012, NGC 4396, NGC 4451, and UGC 11891. This had no material effect on the rotation curve. UGC 3371 also required masking, but for a larger region of 23 pixels. NGC 2644 had 15 pixels that met the criteria for masking; however, they are located toward the center of the galaxy, so their removal would mean losing a significant part of our data. Upon inspection, we chose to leave them in.

\begin{figure*}
\centering
\includegraphics[width=1.0\textwidth]{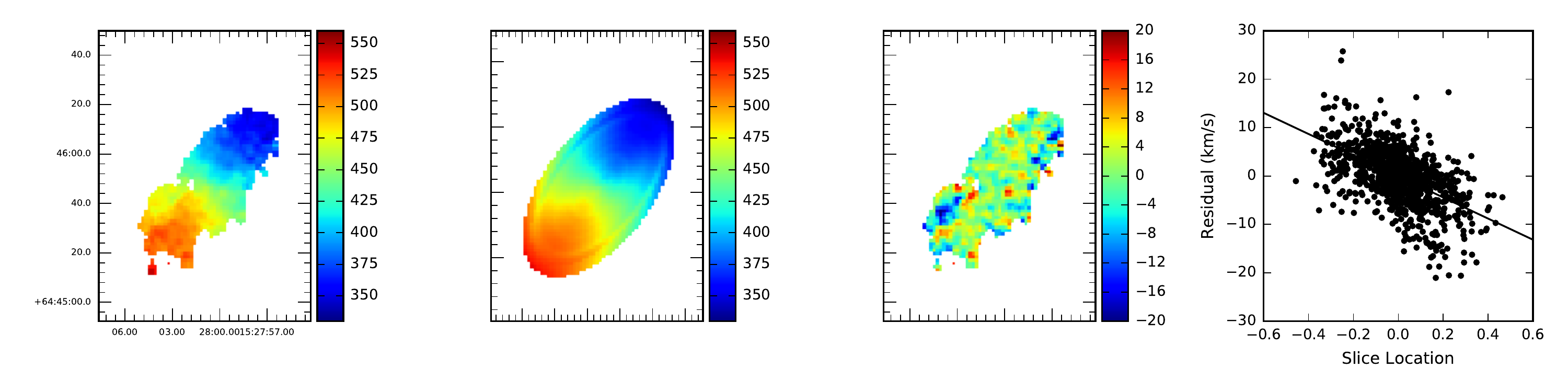}
\includegraphics[width=1.0\textwidth]{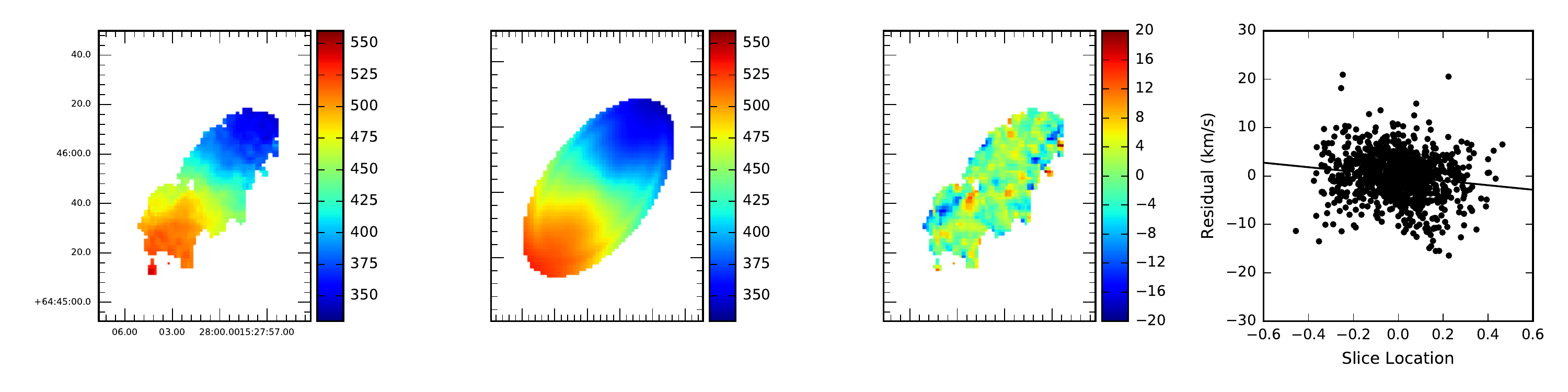}
\caption{Comparison showing the effects of correcting for slice illumination. Each row contains (from left to right) the H$\alpha$ velocity field, DiskFit model, residual field, and plot of residuals against slice location (where within the slice width the H$\alpha$ illumination occurred) for NGC 5949. The top row is before the correction for slit illumination, while the bottom row is after.\label{velc_fig}}
\end{figure*}

\subsection{Slit Illumination Corrections}\label{velc}
In any image slicing IFU, the measured velocity will depend at some level on the distribution of H$\alpha$ within the slice. When using PCWI, the seeing is often smaller than the slice width, so the emission does not illuminate the entire slice. This is potentially problematic because we would measure a different velocity for a point source at the bottom of the slice versus the top. PCWI images have a pixel scale of $0\farcs58$ and a dispersion of 0.32 $\mathrm{\AA}$ per pixel, so we expect the difference in H$\alpha$ velocity across a $2\farcs6$ slice to be $\Delta V = 2\farcs6/0\farcs58 \times 0.32\mathrm{\AA}/6563\mathrm{\AA} \times c \approx 67$ km s$^{-1}$ for a perfect point source.  

In practice, the discrepancy is not as severe as the $\Delta V$ given above. In addition to the diffuse nature of H$\alpha$ reducing the magnitude of this effect, the seeing helps blur out the target over a larger portion of the slice. Furthermore, the effects are still less pronounced in a rotation curve, where velocities are averaged over annuli. Even so, the effect is potentially significant, particularly for our LSB galaxies, whose \ion{H}{2} regions are more point-like, so we chose to model and correct for this effect. 

To correct for this effect, we first need to determine where the emission arrived within each slice. We convolve the narrowband H$\alpha$ images to match the (generally worse) seeing we had with PCWI, and then we project the PCWI slice array onto the image. 

In each spaxel, we compute the centroid of the H$\alpha$ emission within the $2\farcs6$ slice width. We expect this position to correlate with a velocity shift, which we can test by examining the correlation with residuals from the DiskFit model.  To determine the correct offsets, we took the DiskFit models made using the uncorrected velocity field and plotted the residuals against slice location. The top row of Figure \ref{velc_fig} shows the original data and the steps of this process for NGC 5949. As expected, there is a clear correlation between the position in the slice and the residual. 

We use the slope of this correlation to estimate the correction in velocity as a function of position. We repeated the data reduction process using this correction to shift the velocities. The resulting corrected velocity field, DiskFit model, residual field, and residual plot can be seen in the bottom row of Figure \ref{velc_fig}. The differences in the velocity fields are subtle, but the residual correlation plot shows that the correction did remove the dependence on slice location.

This procedure was done for all of the galaxies in our sample, and in all cases using the corrected field improved the $\chi^2$ value of the DiskFit output. As mentioned, the change to the velocity field is subtle, and consequently the correction generally did not make a significant change to the rotation curve, though it often decreased the errors. The maximum difference in velocity ranges from less than 1 to 15 km s$^{-1}$, with a median of 3.4 km s$^{-1}$. The typical velocity correction is thus comparable to the motions of the ISM (6 km s$^{-1}$) included in our DiskFit model. 

The top panels in the Appendix figures show the rotation curves from the corrected fields and their errors in bold, while the rotation curves from the uncorrected fields (No Illum. Corr.) are shown with reduced opacity. For the remainder of our work, we will use the fields corrected for slit illumination. 

\subsection{Bisymmetric Noncircular Motions}\label{bar}
While our intent was to select galaxies with no obvious central bar, it is still possible that some of our galaxies might have noncircular motions induced by a bar that are dynamically significant. In particular, NGC 2976 \citep{Menendez, SandS, Valenzuela} and NGC 6503 \citep{Kuzio} have previously been suggested to contain bars. To investigate this possibility, we used DiskFit to fit for bisymmetric noncircular motions in each galaxy (see Equation \ref{bisym-eq}) and compared the results to the rotation curves previously derived from the radial flow model. In most cases, there was no appreciable difference between the two. For six galaxies, the fit did not converge and the new rotation curves were oscillatory and had enormous errors. If a galaxy fell into either of these cases, we discarded the nonaxisymmetric model. The remaining four galaxies (NGC 949, NGC 2976, NGC 3622, and NGC 4376) showed reasonable variation and could plausibly contain a dynamically significant bar. 

The top panels in the Appendix figures show the bisymmetric rotation curves in purple for the four galaxies listed. NGC 949 shows good agreement in the inner part but shows a drop in the outer parts of the curve that is not seen in the radial flow model. While the shape and magnitude of the different curves for NGC 2976 are nearly the same, the behavior in the center is different. As this region is important to our mass modeling, we chose to continue to consider the bisymmetric case. NGC 3622 and NGC 4376 both show larger velocities in the outer regions of the curve for the bisymmetric fit, and NGC 3622 also disagrees in the center. 

Upon closer inspection, for NGC 3622 the residuals between the H$\alpha$ velocity field and the radial flow model show a distinct central distortion. This leads us to conclude that NGC 3622 is a clear case for containing a central bar, so we will use the bisymmetric model as our fiducial case for this galaxy. We were not able to come to a clear conclusion for the remaining three galaxies. For this reason, for all four galaxies discussed above, we will consider both the radial flow and bisymmetric rotation curves in our analysis and compare results in the mass-model stage of Paper II.

\subsection{Covariance Matrix for Error Estimation}
DiskFit uses a bootstrap technique to measure the scatter in the model parameters about their optimal values. A bootstrap sample is generated by adding randomly drawn residuals (from the distribution with the minimum $\chi^2$ value) to the optimal velocity field model \citep{SandS}. The minimization is done again using the bootstrap velocities, and this process is repeated 500 times. The standard deviation of each parameter from all bootstrap realizations is then given as the uncertainty. 

DiskFit provides a log containing a list of the best-fitting model parameters determined at each bootstrap realization. For the $N$ radial bins in each galaxy, we will use these logged data to compute the $N \times N$ covariance matrix for use in the mass modeling contained in Paper II. Correlations between radial bins could easily arise in our data, due to dependence on parameters like inclination or PA. Using the covariance matrix is preferable because it takes such correlations into account.

\section{Comparison to Molecular Gas Observations}\label{CO_comp}
As discussed in Section \ref{sample}, this work is being conducted alongside a similar investigation using CO observations from CARMA. Of the galaxies with good CO detections in \cite{Mai1}, we overlap with 11. Here we compare the resulting velocity fields for this subsample of galaxies, while we will compare the derived dark matter distributions in Paper 2. A similar comparison has been done between \ion{H}{1} and CO by \cite{Frank}.

Before making our comparison, we first need to mask the CO velocity fields. The CO observations suffer from some localized fluctuations and low signal-to-noise pixels.  As we want to examine the global dynamics, we filtered out pixels that had a larger than 50 km s$^{-1}$ difference from the median of 25 surrounding pixels, as well as any isolated single pixels. 

For each of the 11 galaxies, Figure \ref{fig:COvHa} shows the H$\alpha$ velocity field (which has been convolved to match the resolution of the CARMA data, typically $\sim$3\farcs5) in the left column, the masked CO velocity field in the middle column, and the difference between the fields in the right column. The standard deviation of each difference field, $\sigma_{dv}$, is also given in the right panel. We find good agreement between the two kinematic tracers, finding a median of $\sigma_{dv} = 6.9$ km s$^{-1}$. This is on the order of the turbulent motions of the ISM, which are around 6 km s$^{-1}$, as discussed in Section \ref{ISM}. 

\begin{figure*}
\centering
\includegraphics[width=1.0\textwidth]{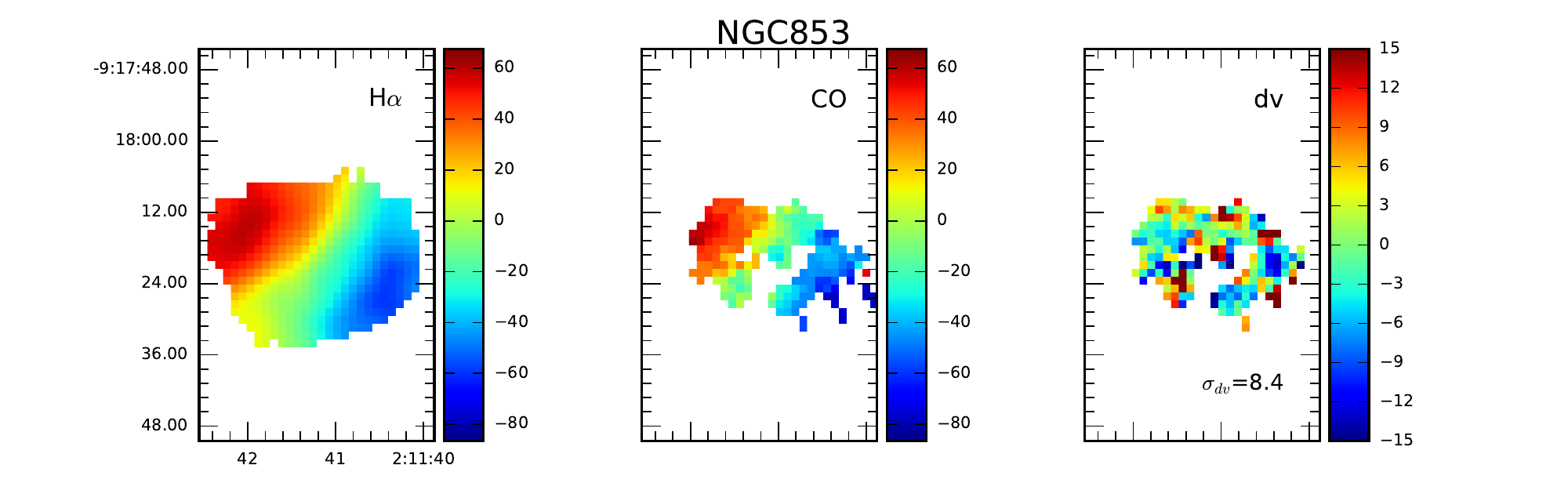}
\includegraphics[width=1.0\textwidth]{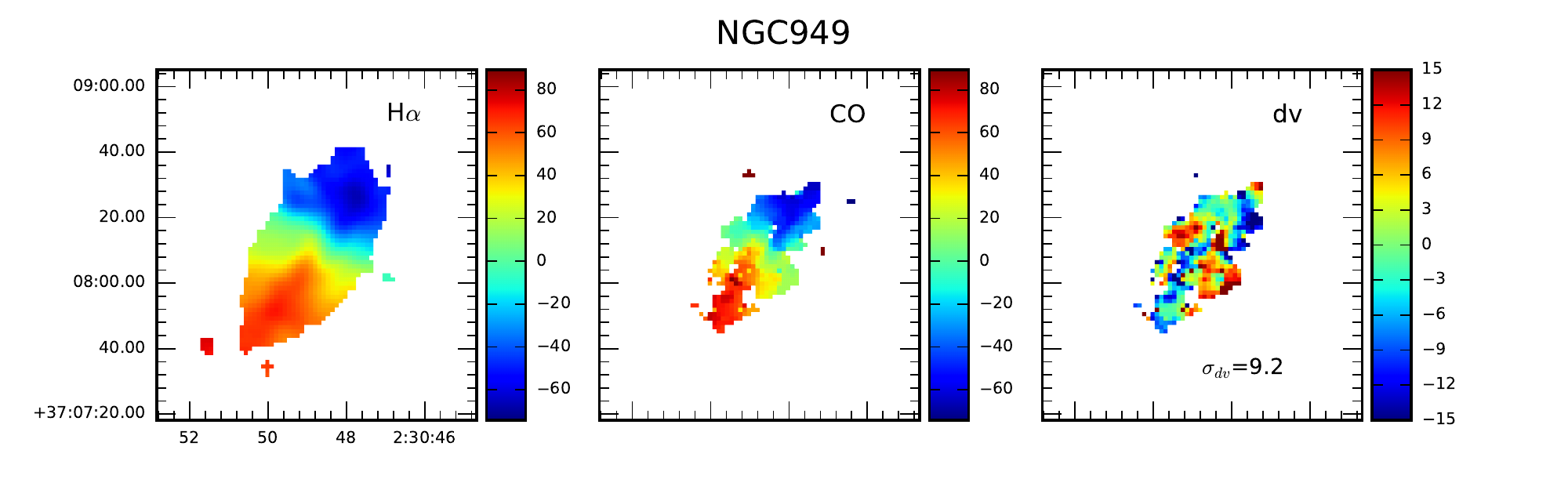}
\includegraphics[width=1.0\textwidth]{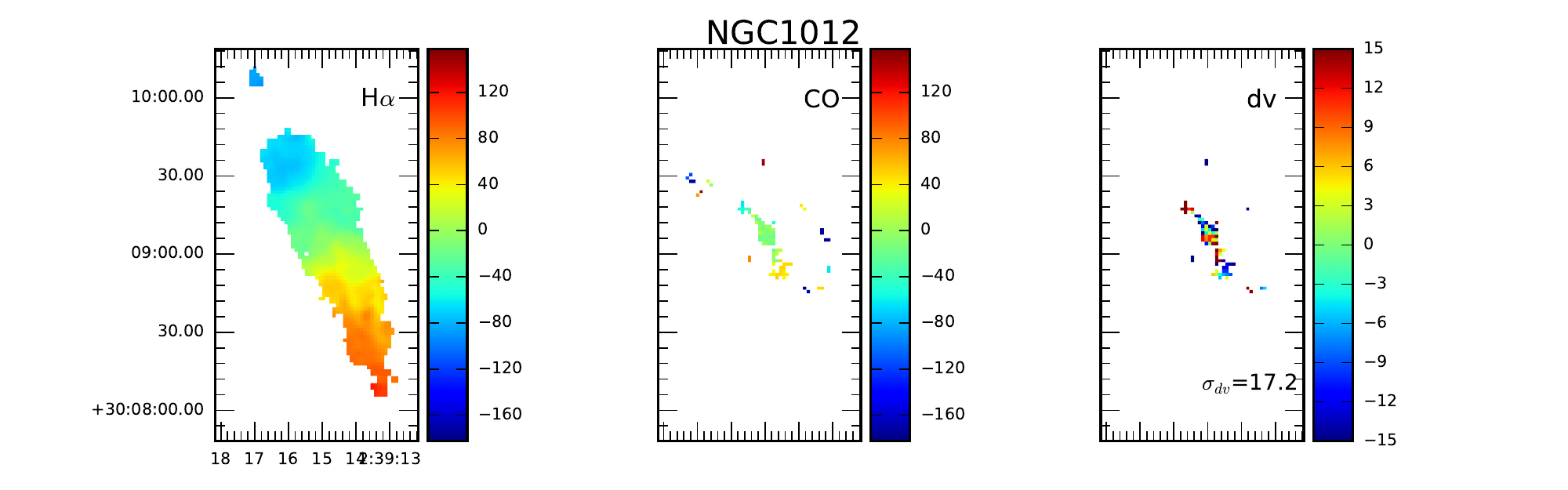}
\includegraphics[width=1.0\textwidth]{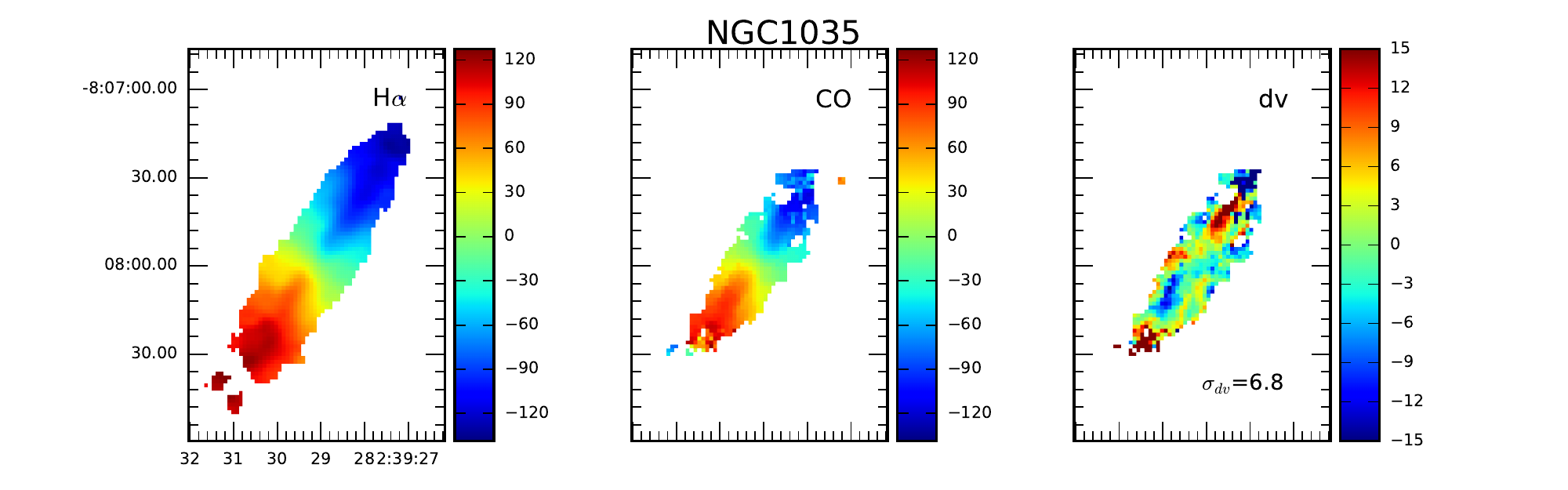}
\end{figure*}
\begin{figure*}
\centering
\includegraphics[width=1.0\textwidth]{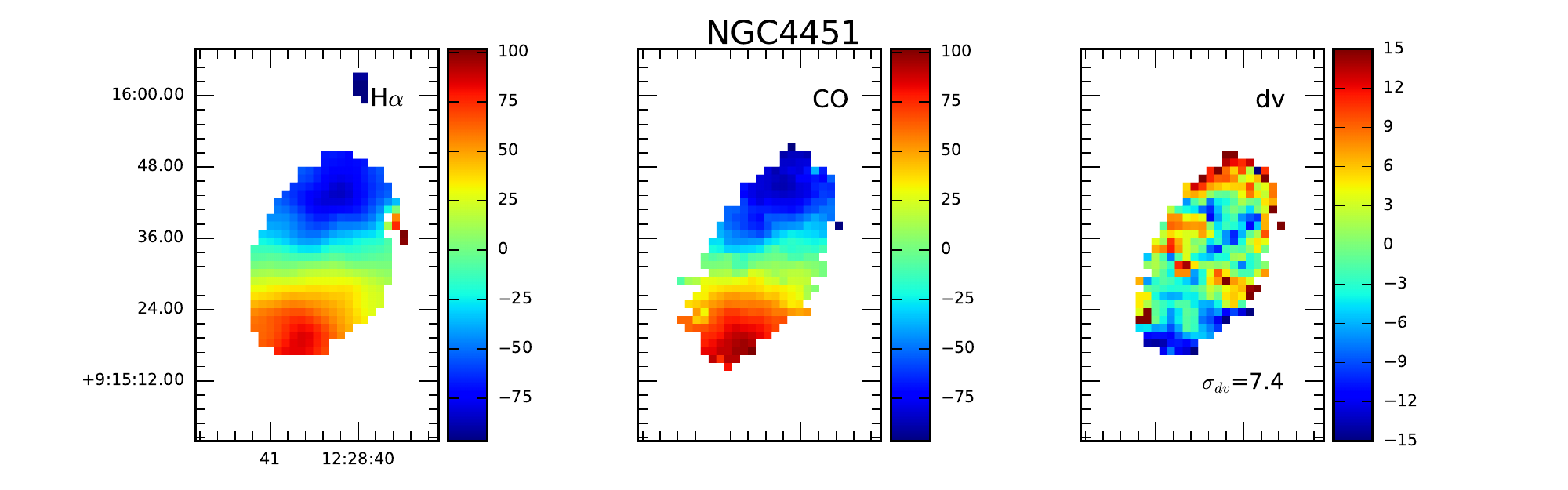}
\includegraphics[width=1.0\textwidth]{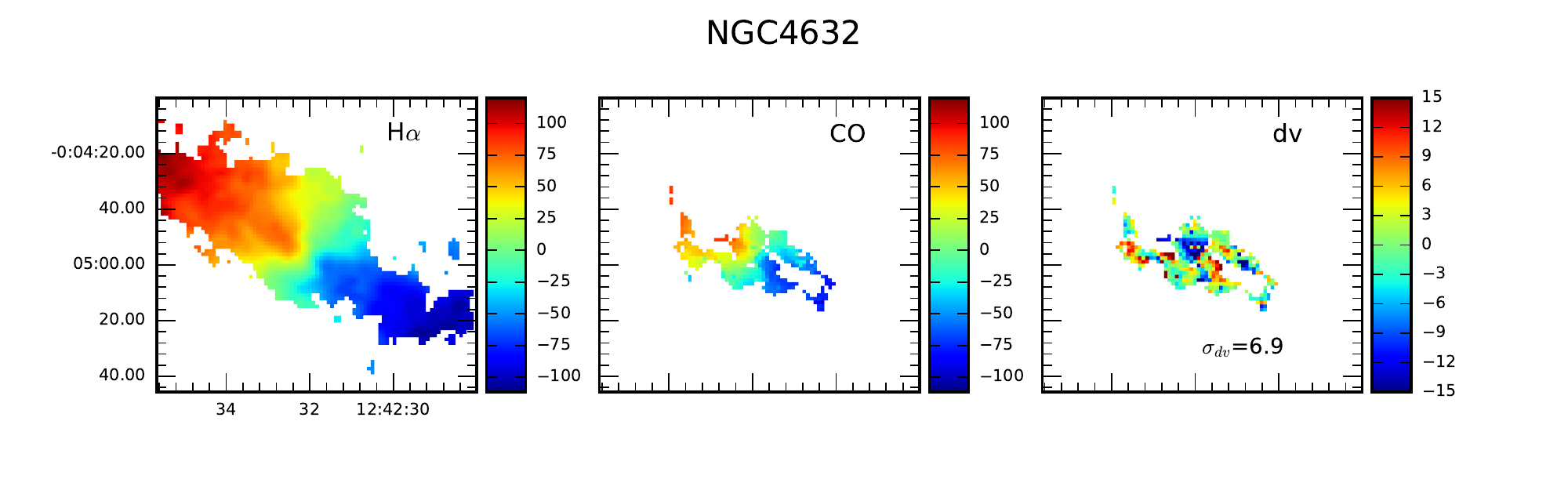}
\includegraphics[width=1.0\textwidth]{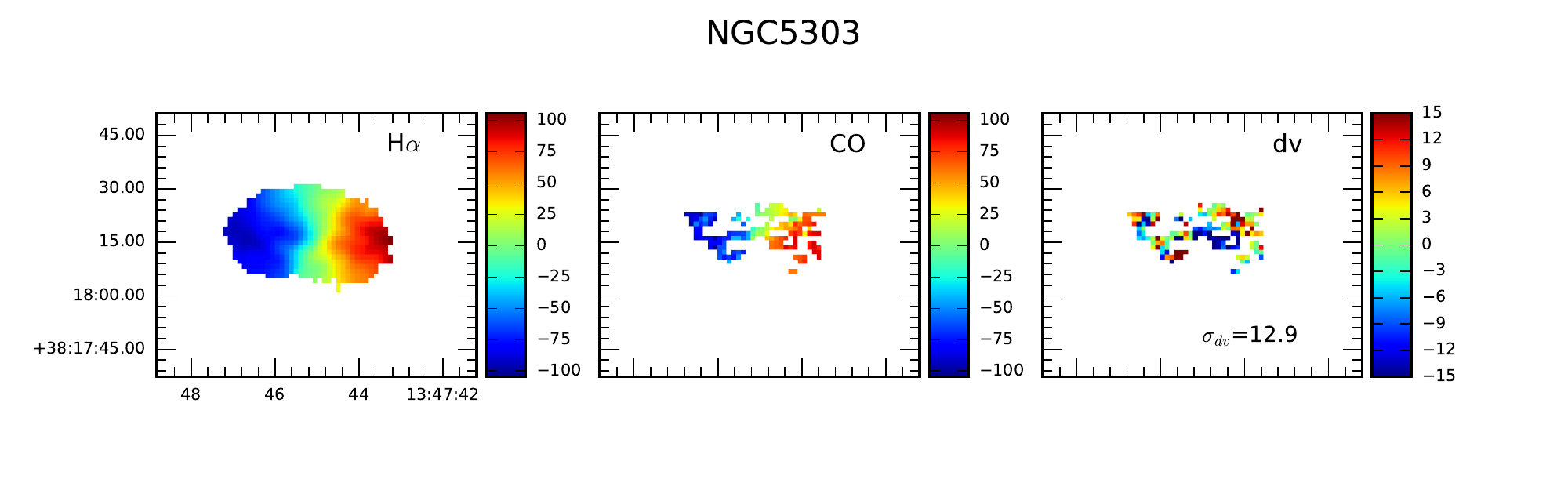}
\includegraphics[width=1.0\textwidth]{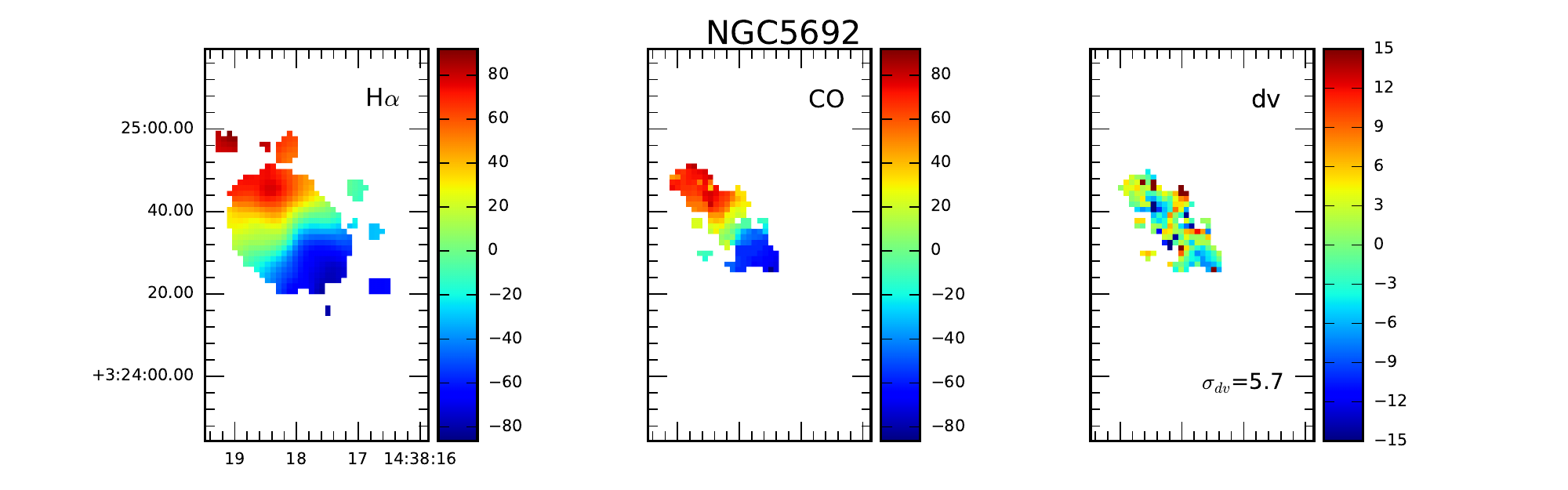}
\end{figure*}
\begin{figure*}
\centering
\includegraphics[width=1.0\textwidth]{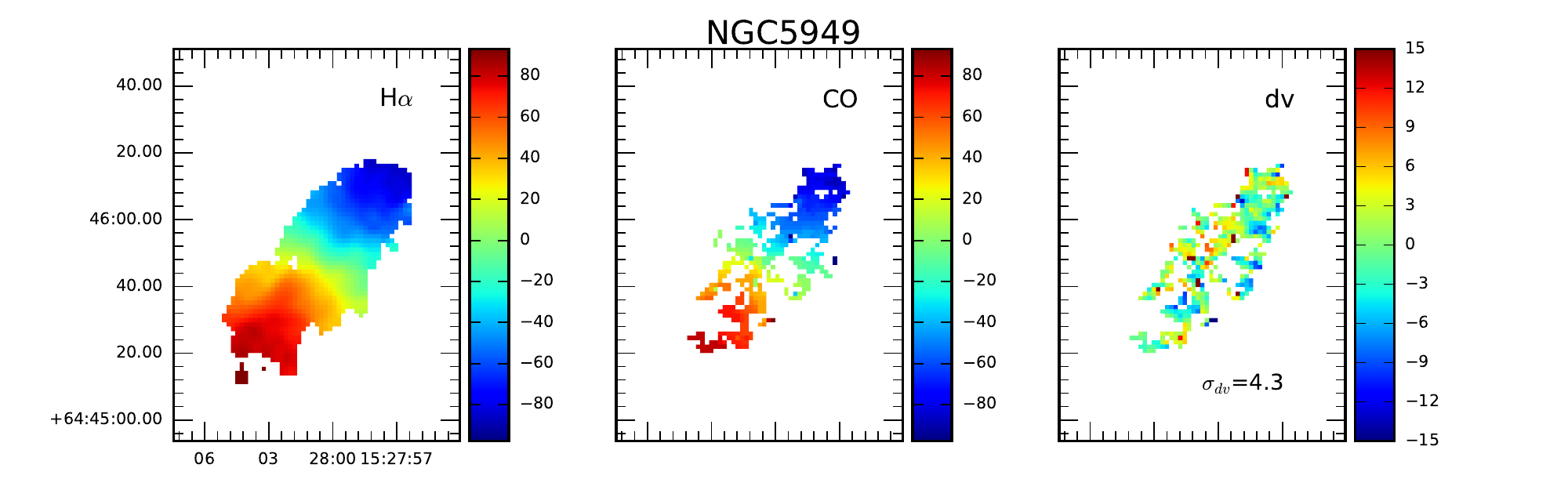}
\includegraphics[width=1.0\textwidth]{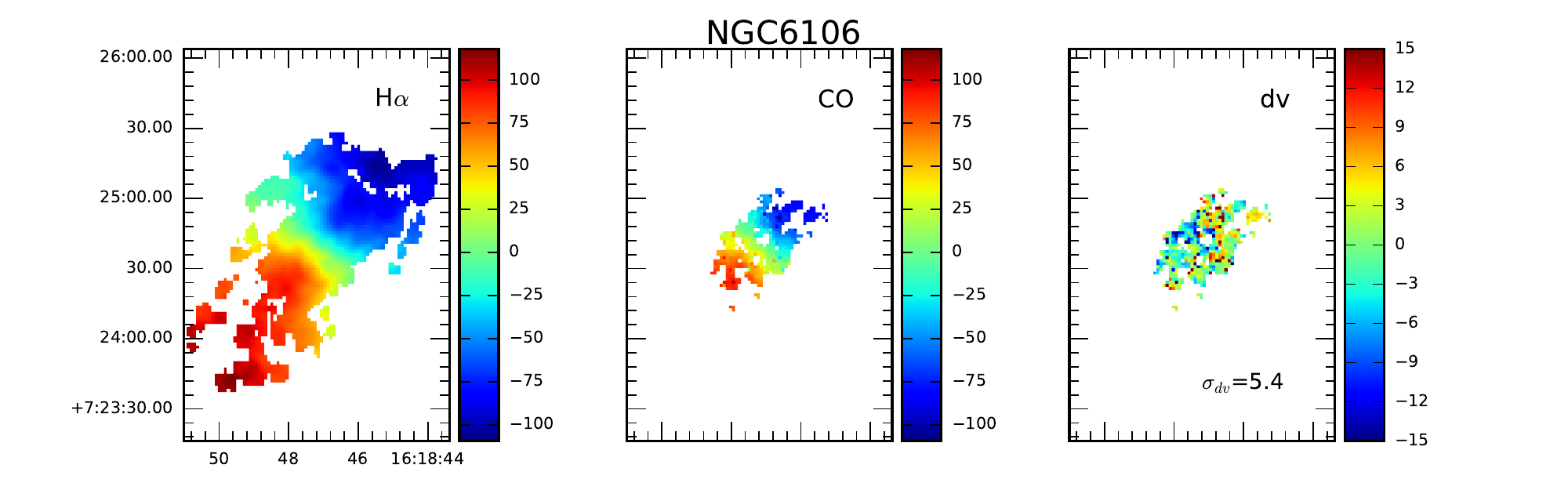}
\includegraphics[width=1.0\textwidth]{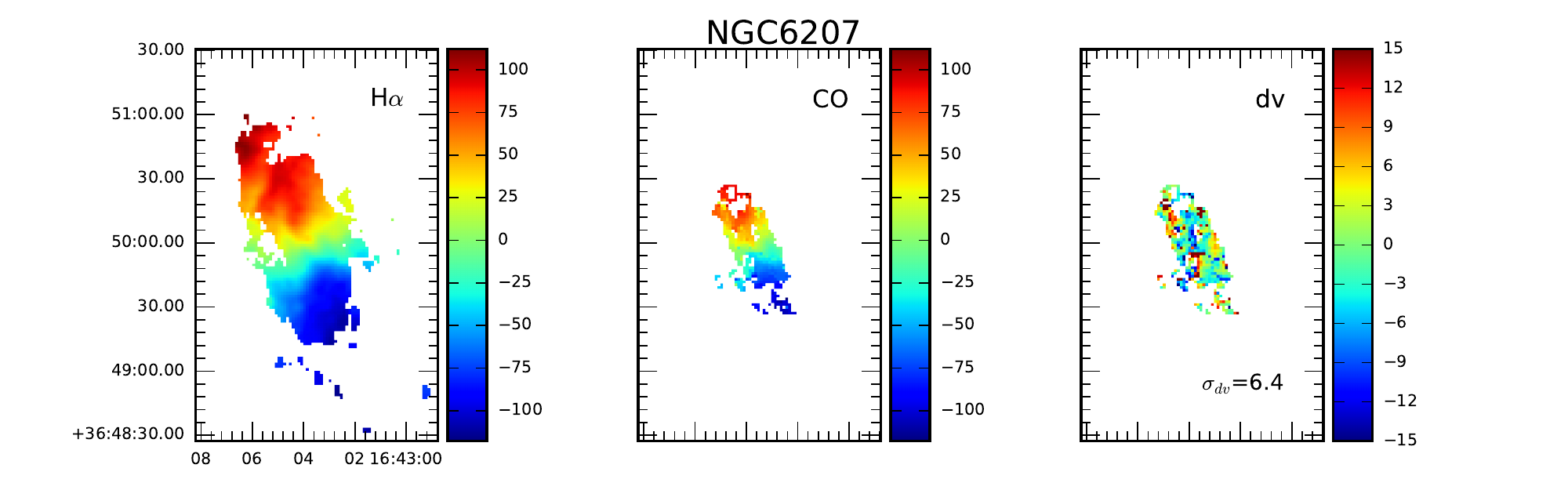}
\caption{From left to right: the H$\alpha$ velocity field (convolved to match the resolution of the CARMA data), the CO velocity field, and the difference between the velocity fields. The standard deviation of the difference is given in the right panels.}
\label{fig:COvHa}
\end{figure*}

NGC 1012 has the largest $\sigma_{dv}$, though the poor agreement appears to be due to a handful of noisy pixels in the CO data rather than a systematic disagreement between datasets. Note that the H$\alpha$ measurements extend to larger radii than those from CO, which will be helpful in constraining the overall shape of the rotation curves, as was demonstrated by \cite{Mai1}. The dispersion of points with large differences is primarily random, though some coherent patches exist. As the rotation curves are derived by averaging over annuli, these patches will not have a significant effect on the rotation curve. 

\begin{figure}\label{COhist}
\includegraphics[width=\columnwidth]{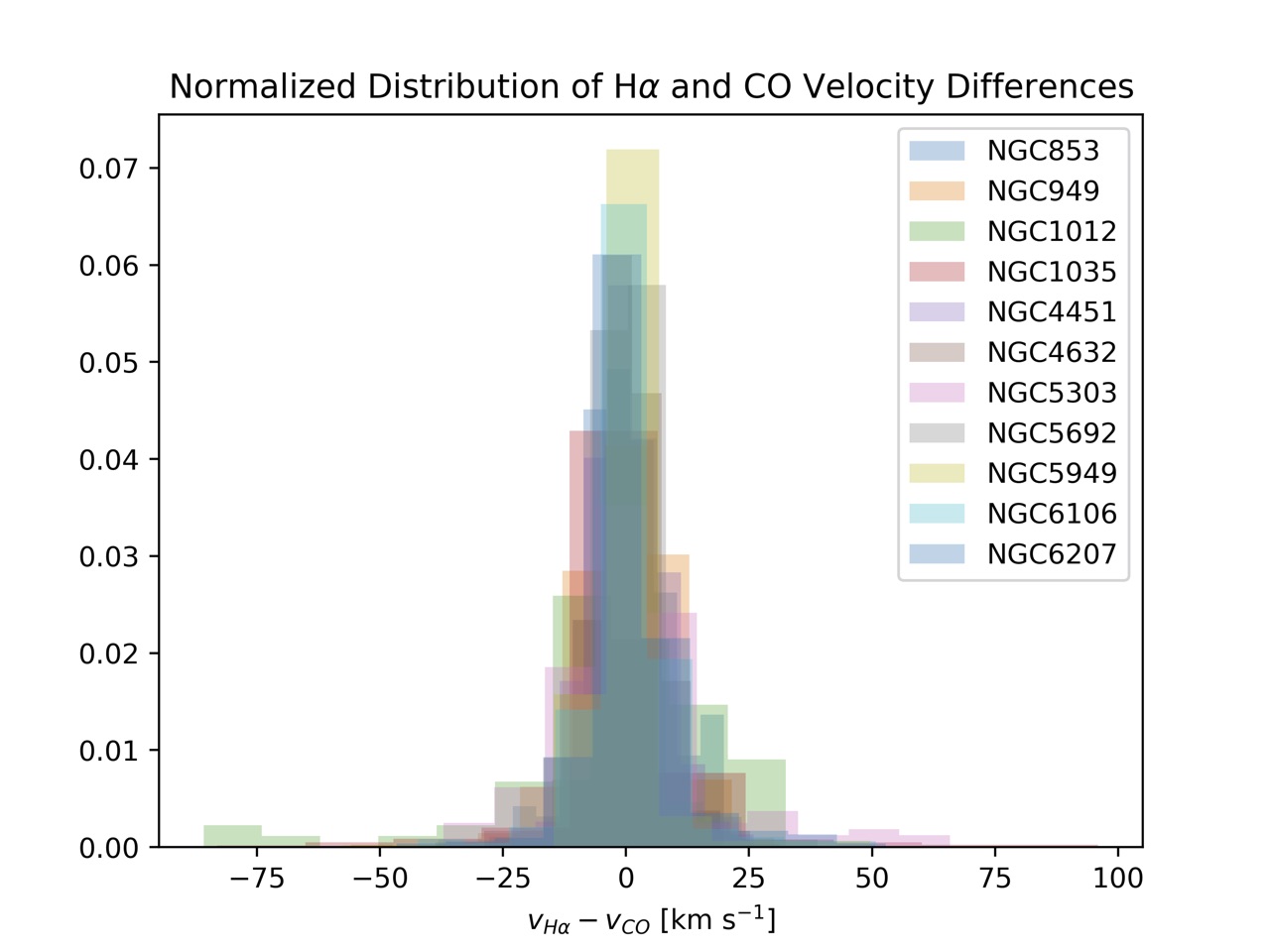}
\caption{Normalized histograms showing the differences between the H$\alpha$ and CO velocity fields for the 11 galaxies in both samples.}
\label{COhist}
\end{figure}

Figure \ref{COhist} contains the normalized histograms of all 11 velocity difference fields (right panels in Figure \ref{fig:COvHa}). Note that for ease of viewing Figure \ref{COhist} has been clipped to only show differences in the range from -100 to 100 km s$^{-1}$. There are a small fraction of outlier pixels in the distributions. It is not clear whether these represent genuine velocity differences between the two gas phases or spurious errors in the maps, but in either case they are sufficiently rare that they do not affect the bulk kinematics.

\section{Literature Comparisons} \label{comp}
Rotation curves derived from H$\alpha$ observations exist in the literature for several of our galaxies. As a verification of our data reduction and fitting procedures, in this section we compare our resulting curves to the previously published ones.  

\begin{figure*}
\centering
\includegraphics[width=\textwidth]{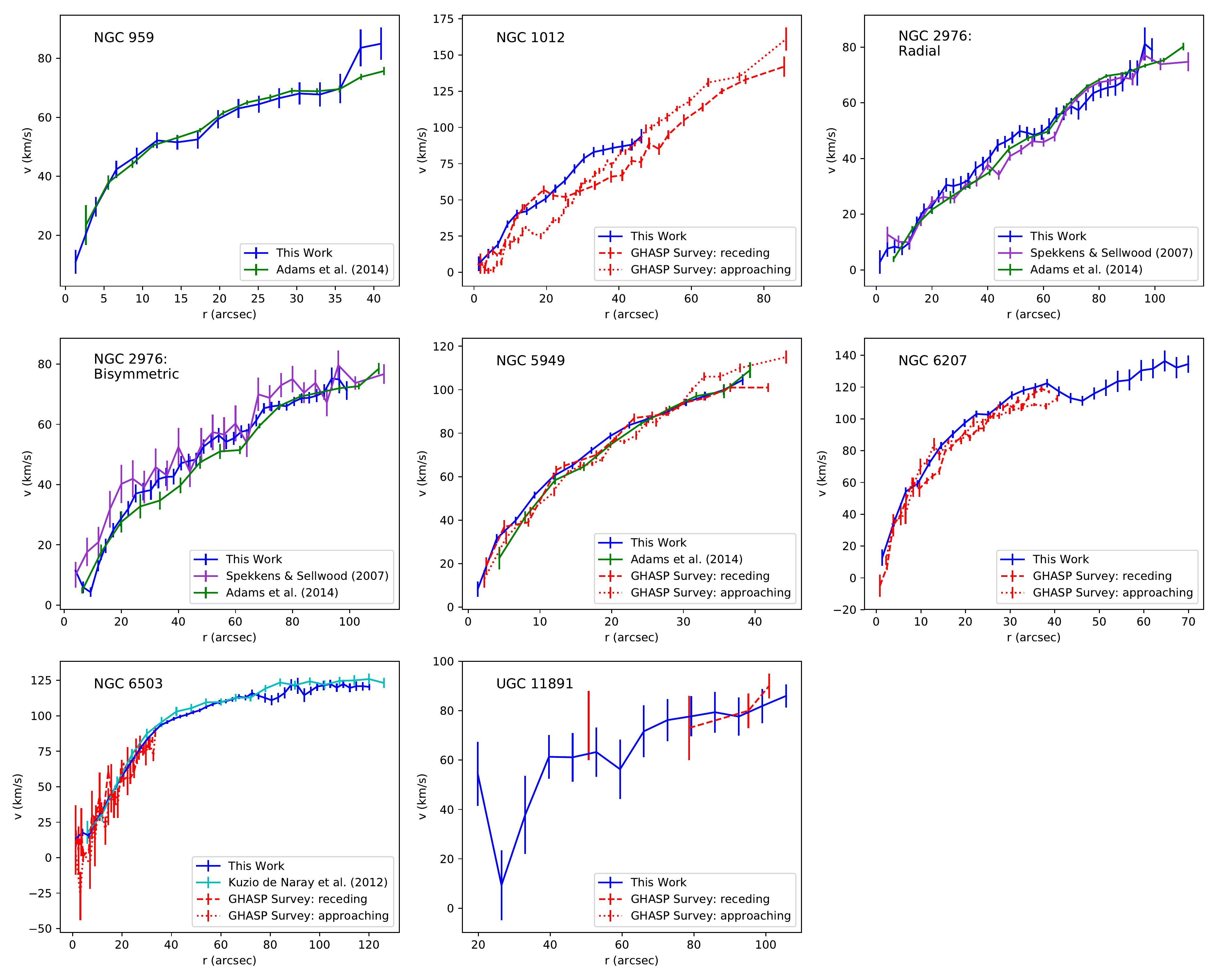}
\caption{Comparisons of our rotation curves with other H$\alpha$ rotation curves in the literature. Our results are typically in good agreement with the existing data. See Section \ref{comp} for further discussion. \label{comparison}}
\end{figure*}

Five of our galaxies (NGC 1012, NGC 5949, NGC 6207, NGC 6503, and UGC 11891) were part of the GHASP Survey \citep{ghasp}. Their data are categorized by whether the velocity came from the approaching or receding side of the galaxy, so for completeness we have plotted the data from both sides in Figure \ref{comparison}. 

The sample of dwarf galaxies studied in \cite{Adams} has three that overlap with our sample, NGC 959, NGC 2976, and NGC 5949. In the case of NGC 2976, we also have rotation curves from \cite{SandS}, who used DiskFit to model the velocity field obtained by \cite{simon} and investigated whether a bisymmetric fit is more appropriate for this galaxy. As discussed in Section \ref{bar}, we also considered this case, as did \cite{Adams}; thus, we are able to compare both radial and bisymmetric options across three sets of independent measurements. Note that we use the gas-traced models from \cite{Adams}, although they additionally provided stellar-traced models. 

In addition to the GHASP Survey, NGC 6503 has been studied in detail by \cite{Kuzio} using H$\alpha$, CO, and \ion{H}{2} velocity fields. They also applied DiskFit to H$\alpha$ data, and while they considered a model with a bisymmetric distortion, they concluded that a rotation-only model is sufficient. As we came to a similar conclusion, we will only compare our radial flow fits. 

All plots comparing the various rotation curves can be seen in Figure \ref{comparison}.

\begin{enumerate}
\item{NGC 959 agrees very well with the \cite{Adams} rotation curve, only deviating notably at the outer edge of the galaxy. This region will not impact the inner slope of the dark matter density profile. }

\item{NGC 1012 has some deviation from the GHASP data, particularly in the middle of the galaxy, although the overall shape is similar. We find a larger rotation velocity by 10-15 km s$^{-1}$ out to $r = 40\arcsec$. }

\item{NGC 2976 has generally good agreement with \cite{Adams} and \cite{SandS} for both models, with a few exceptions. In the radial flow model, the behavior at the innermost points is inconsistent across all three datasets, a potentially important difference due to its location, although the error bars imply the difference in this region is not very statistically significant. The problem persists in the bisymmetric plot, where the overall differences between datasets are somewhat more pronounced. }

\item{NGC 5949 aligns with the GHASP data well, as do the \cite{Adams} data in the outer parts, although for the rest of the curve our rotation velocities are $\sim$5 km s$^{-1}$ larger. }

\item{NGC 6207 also agrees well with the GHASP data.}

\item{NGC 6503 is very similar to the \cite{Kuzio} data; the variations are small and primarily in the outer part of the curve. The GHASP data are noisy but also in agreement. }

\item{UGC 11891 only has a few data points from the GHASP observations to compare, as they could not measure the inner part of the rotation curve, but we agree on the velocity at large radii. }
\end{enumerate}

Our general agreement with external datasets is good. The small differences we see are often in normalization, which can simply be due to small differences in the inclination angle, while the overall shape and slopes tend to agree.

\section{Summary and Further Work}
We have presented high-resolution 2D H$\alpha$ velocity fields obtained with PCWI for a sample of 26 low-mass galaxies spanning the range $\log L_r/L_{\sun}= 8.4-9.8$, $v_{\rm max} = 50-140$ km s$^{-1}$. This is among the largest surveys presenting rotation curves of dwarf galaxies. Due to the sensitivity of PCWI, we typically detect continuous H$\alpha$ emission throughout the disk, which enables us to trace the rotational motion and detect noncircular motions. The H$\alpha$ rotation curves generally show smooth behavior. They typically extend over most of the optical disk, and they sometimes reach the flat part of the rotation curve (see the Appendix). The amplitude of noncircular motions is also usually small, implying that most of our sample will be suitable for measuring robust dark matter profiles. In some cases, there is evidence of substantial noncircular motions that are a non-negligible fraction of the tangential motion over a significant part of the disk (e.g., NGC 746, NGC 949, NGC 3622, and UGC 01104). In Paper II, we will assess the suitability of each galaxy for mass modeling before considering their dark matter distributions. 

We varied several aspects of the analysis to test the robustness of the derived rotation curves. In most cases, the rotation curves are insensitive to whether the input disk geometry is taken from images or kinematic data. For a third of the sample, however, the values from the kinematic fit are clearly inconsistent with the photometric values. We fixed the center, PA, and ellipticity to the photometric measurements when this was the case. We also investigated the possibility of bisymmetric distortions in the velocity fields that might arise from weak bars. In most cases, the rotation curves derived using a bisymmetric model either were indistinguishable from a simple radial flow model or failed to converge, presumably because the model cannot describe the data well. Only for one galaxy (NGC 3622) did the bisymmetric model provide a clearly superior fit, while for three others some small differences between the bisymmetric and radial flow models were apparent. We will retain both sets of rotation curves to analyze in Paper II. 

We presented a method to correct velocity fields obtained using image slicing IFUs for errors that arise when the line emission does not fully illuminate a slice. This method, which requires a narrowband image, could prove useful for analyzing data obtained using other image slicers, including the Cosmic Web Imager located on the Keck II telescope \citep{KCWI}. For our dataset, we found that these corrections tend to average out within annuli and so produce very modest changes to the rotation curves (as seen in the Appendix figures). The difference the correction made to an example velocity field and residual map can be seen in Figure \ref{velc_fig}. 

Comparisons of the kinematics of dwarf galaxies measured using different tracers are still relatively rare. A detailed comparison of H$\alpha$ and CO kinematics has so far been presented in the literature for only two galaxies in our mass range, NGC 2976 \citep{simon} and NGC 6503 \citep{Kuzio} (for a comparison of more massive galaxies, see \citealt{E-C}). A possible cause for concern is that outflows from \ion{H}{2} regions could induce significant nongravitational motions in the H$\alpha$-emitting gas. Additionally, H$\alpha$ emission can extend far from the disk, and emission from this extraplanar gas could potentially cause the measured kinematics to systematically deviate from the motion in the disk plane \citep{OmanHI}. In this paper, we have compared velocity fields obtained from these tracers for 11 galaxies within our sample (Figure \ref{fig:COvHa}). We find that the velocity fields are quite similar, and the rms variation between the two tracers is comparable to the level of ISM turbulence. Therefore, we do not expect significant differences in rotation curves obtained using H$\alpha$ and CO kinematics (as shown by \citealt{Mai1}), or in the inferred dark matter profiles, which we will analyze in Paper II. The consistency of the H$\alpha$ and CO kinematics, coupled with the study of the stellar kinematics of similar galaxies by \cite{Adams}, shows that the concerns mentioned above have little practical effect, at least for late-type galaxies with masses similar to those of our sample. 

In Paper II, we will use the rotation curves derived here to measure the mass distributions in our sample of low-mass galaxies. Using the optical and infrared images presented here, along with the CO maps obtained with CARMA, we will measure the contribution of stellar and molecular gas to the rotation curves, allowing us to isolate the dark matter distribution. Our sample provides a large and well-characterized set of velocity fields with which to study the diversity of dark matter profiles and explore the implications for the cusp-core problem.

\section*{Acknowledgements}
We would like to thank Joshua Adams, Rachel Kuzio de Naray, and Kristine Spekkens for providing the rotation curve data used for comparisons made in Figure \ref{comparison}.

This research has made use of the NASA/IPAC Extragalactic Database (NED), which is operated by the Jet Propulsion Laboratory, California Institute of Technology, under contract with the National Aeronautics and Space Administration.

This research has made use of the SIMBAD database, operated at CDS, Strasbourg, France.

\appendix
In Figures 9-13, we provide images and rotation curves for each of the 26 galaxies in our sample. The top panels show the rotation curves derived from H$\alpha$ kinematics, with the tangential velocities plotted with solid lines and radial velocities plotted with dashed lines. The rotation curves generated with all parameters fixed to photometric values are blue, those with all parameters allowed to vary are in green, and bisymmetric fits are in purple (when relevant). As described in Section \ref{velc}, we correct our measured velocities for H$\alpha$ slit illumination location. The rotation curves made with the uncorrected velocity fields (No Illum. Corr.) are shown behind the corrected ones with reduced opacity. Error bars are only shown on the curve that we will use for our analysis.

The bottom panels provide optical images, H$\alpha$ intensity maps, and derived velocity fields from PCWI. The bottom left panels are $r$-band images of each galaxy, taken in addition to narrowband H$\alpha$ images primarily using SPICAM at APO (see Section \ref{phot} for more details). The bottom middle panels show the H$\alpha$ intensity maps constructed from our PCWI exposures, which have been overlaid with contours from the corresponding narrow-band images. We find good agreement between the contours from the narrowband imager and the PCWI data. The bottom right panels show the final velocity fields that are corrected for slit illumination (see Section \ref{velc}) and have the systemic velocities subtracted.

\begin{figure*}
\centering
\includegraphics[width=0.83\textwidth]{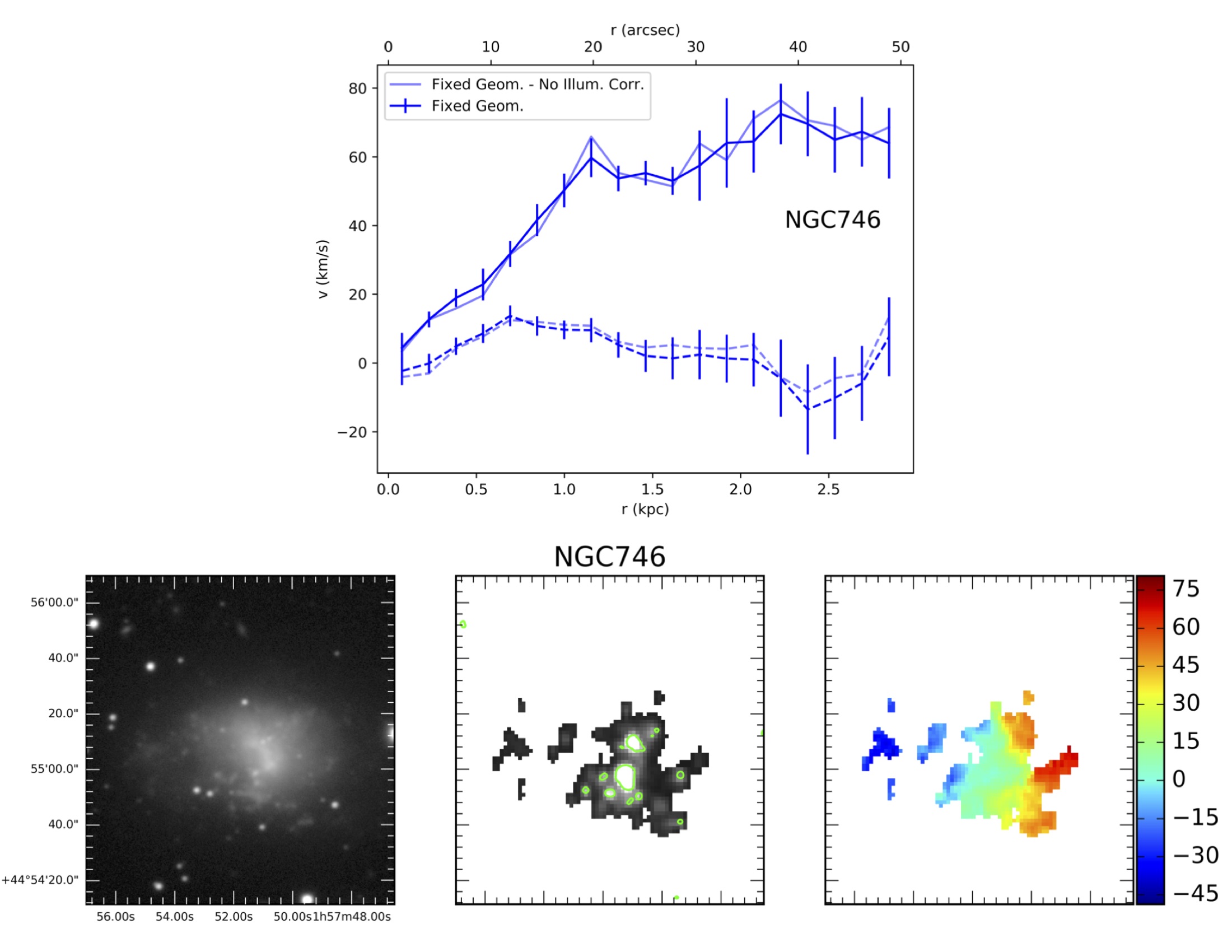}
\includegraphics[width=0.83\textwidth]{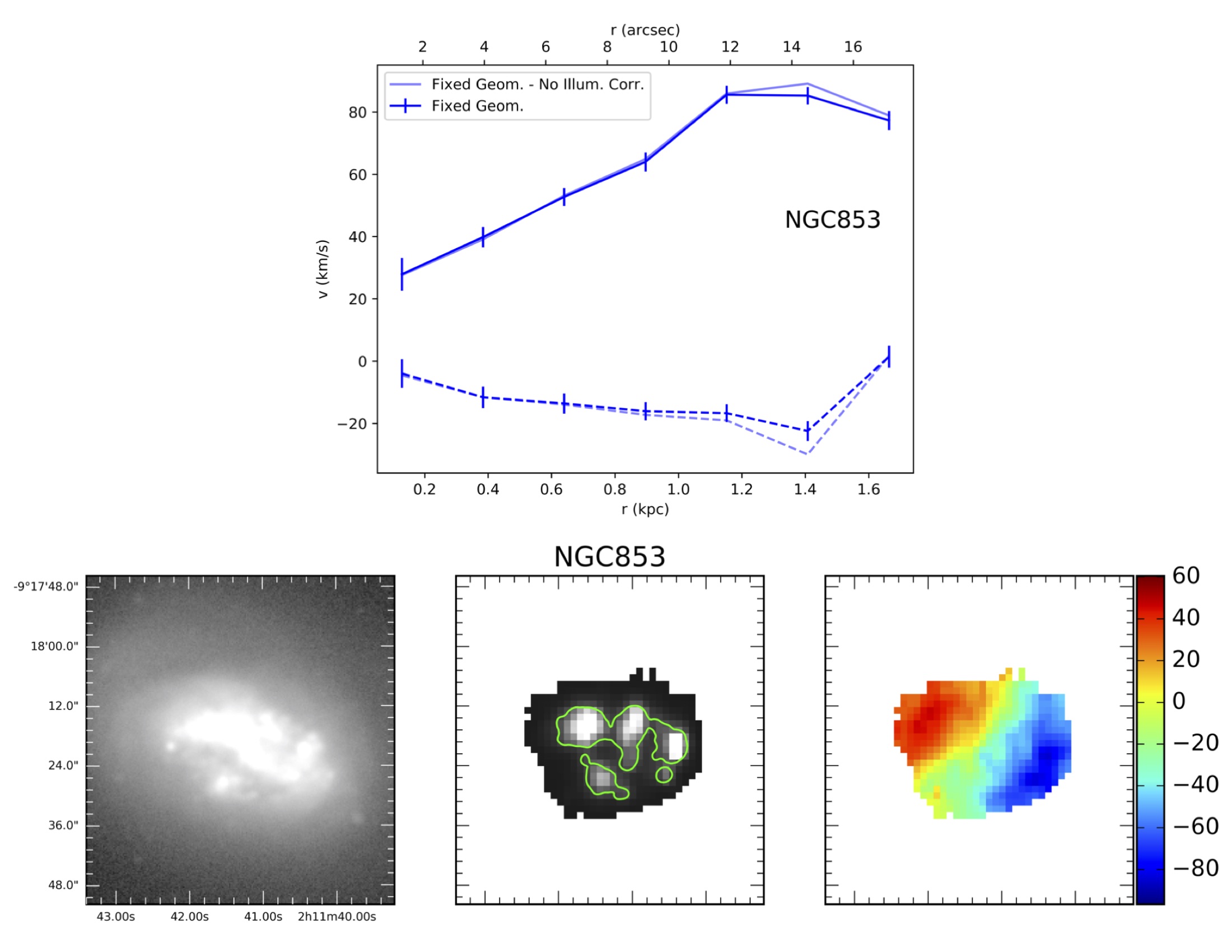}
\end{figure*}
\begin{figure*}
\centering
\includegraphics[width=0.83\textwidth]{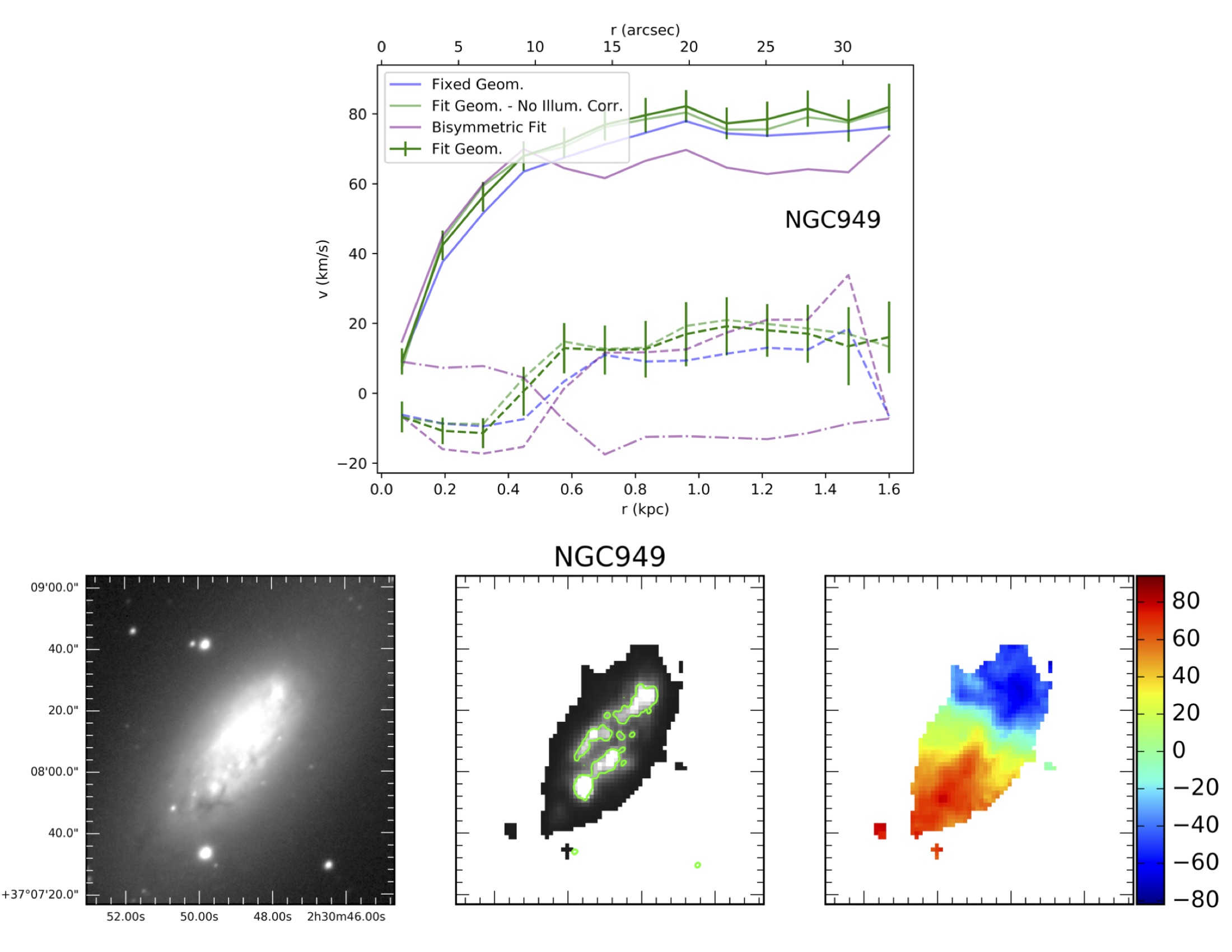}
\includegraphics[width=0.83\textwidth]{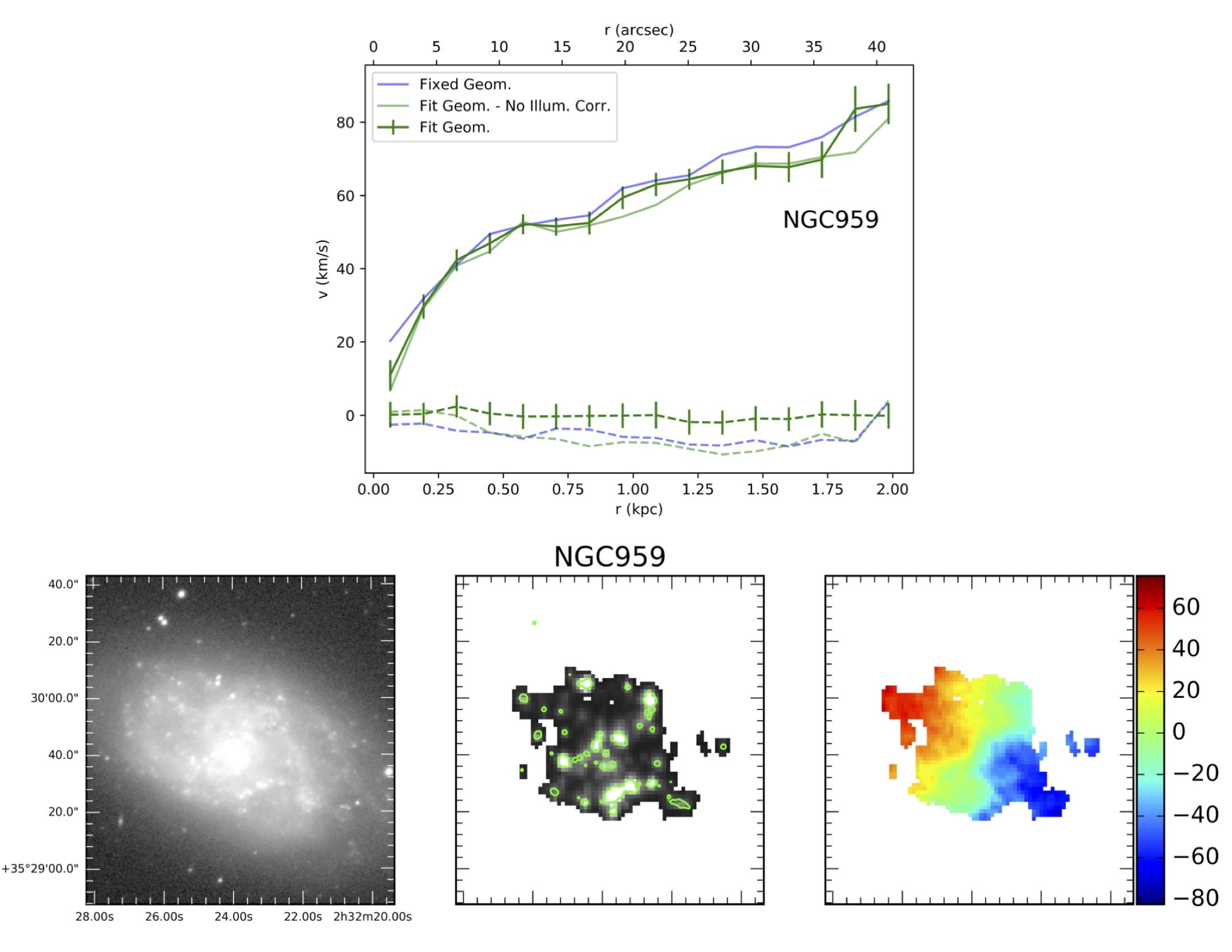}
\end{figure*}
\begin{figure*}
\centering
\includegraphics[width=0.83\textwidth]{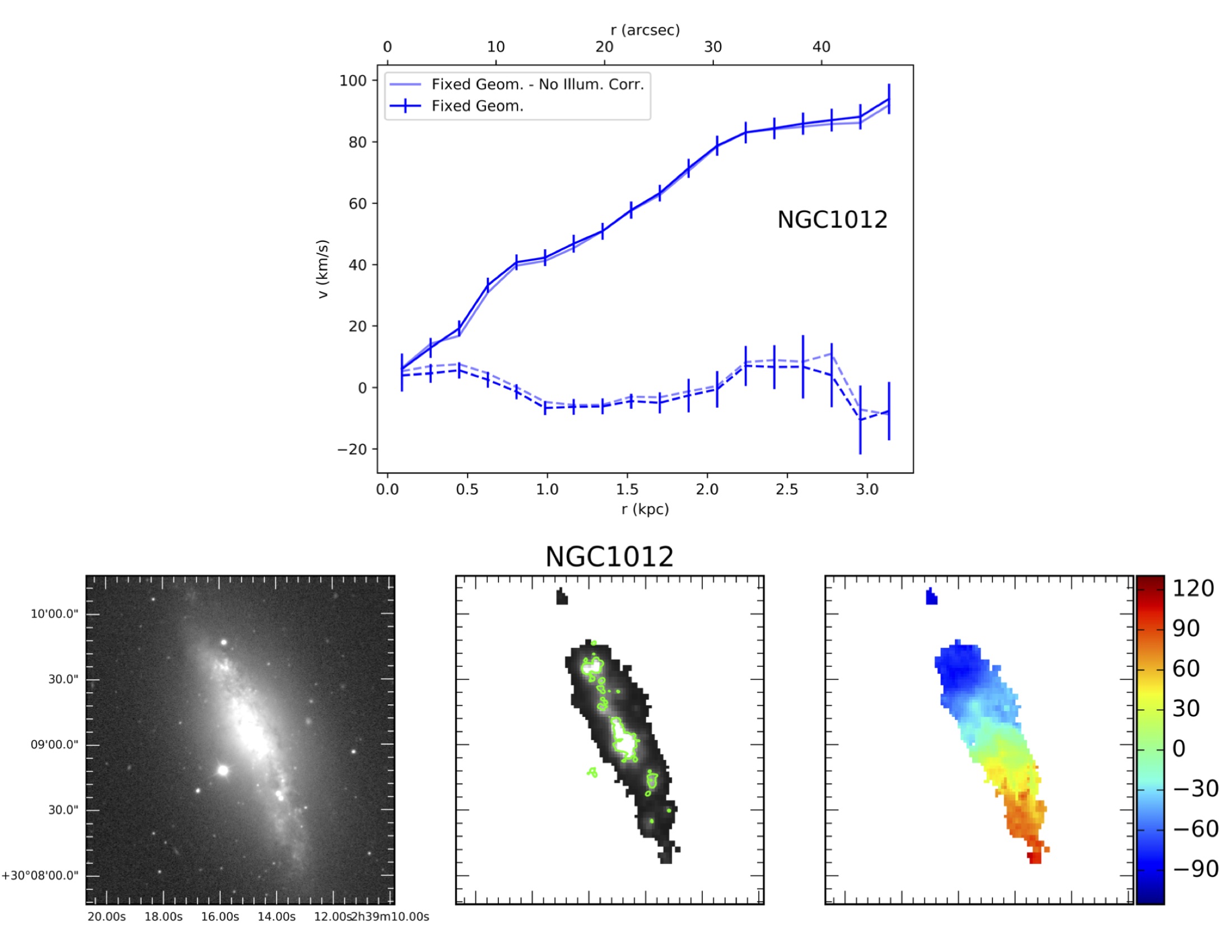}
\includegraphics[width=0.83\textwidth]{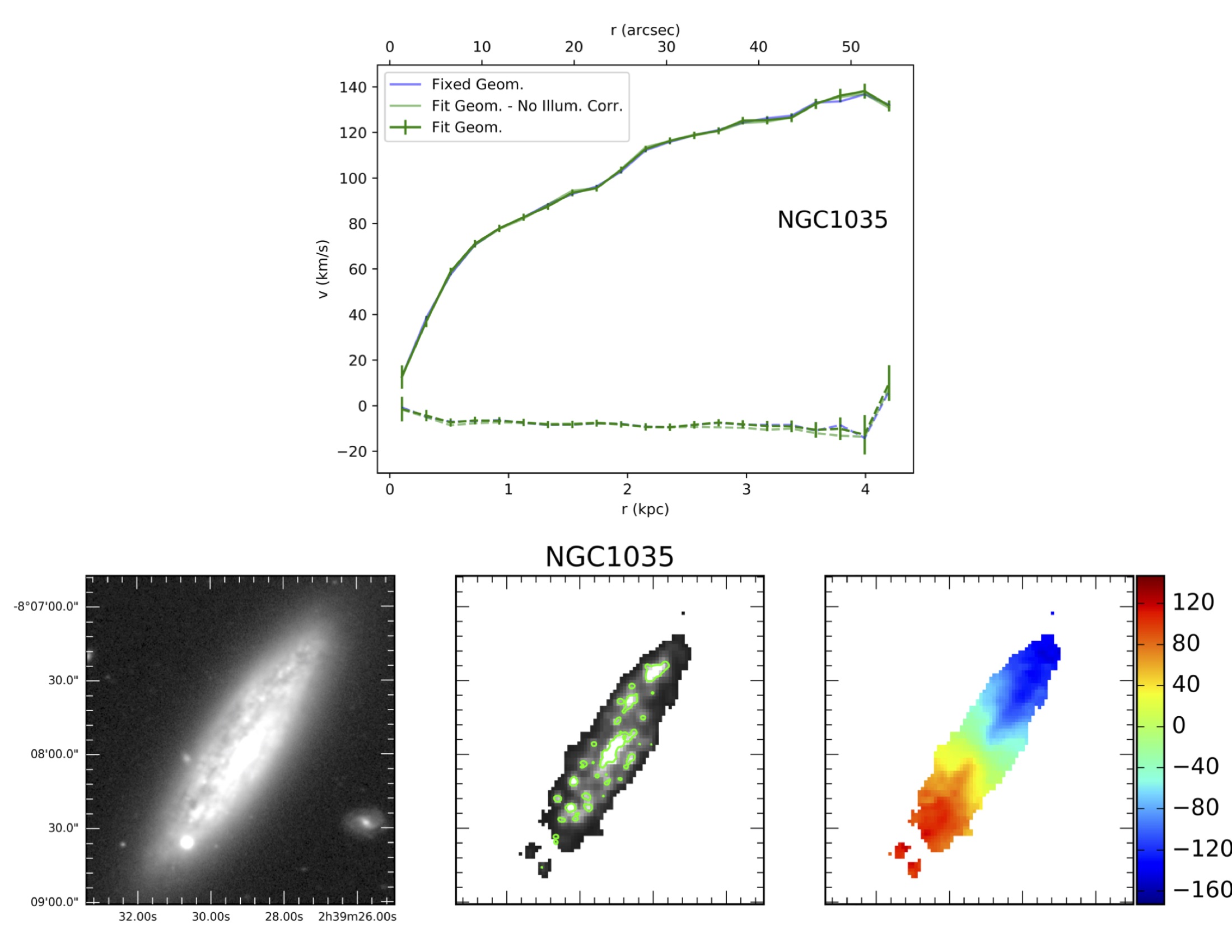}
\caption{Galaxies NGC 746, NGC 853, NGC 949, NGC 959, NGC 1012, and NGC 1035. For each, the top panel shows the H$\alpha$ rotation curve, while the bottom panels (from left to right) show the $r$-band image, total H$\alpha$ flux (with contours from photometry), and velocity field.}
\end{figure*}

\begin{figure*}
\centering
\includegraphics[width=0.83\textwidth]{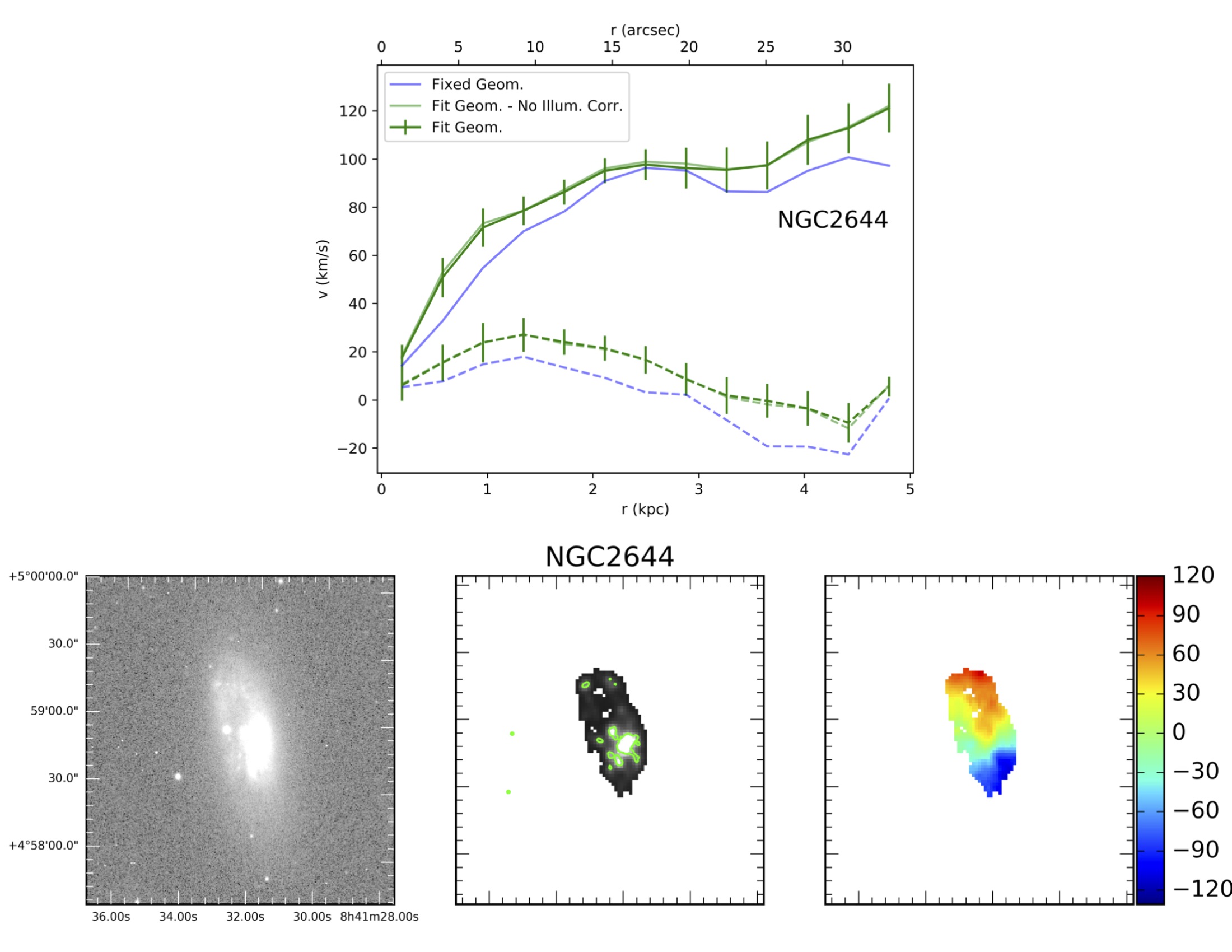}
\includegraphics[width=0.83\textwidth]{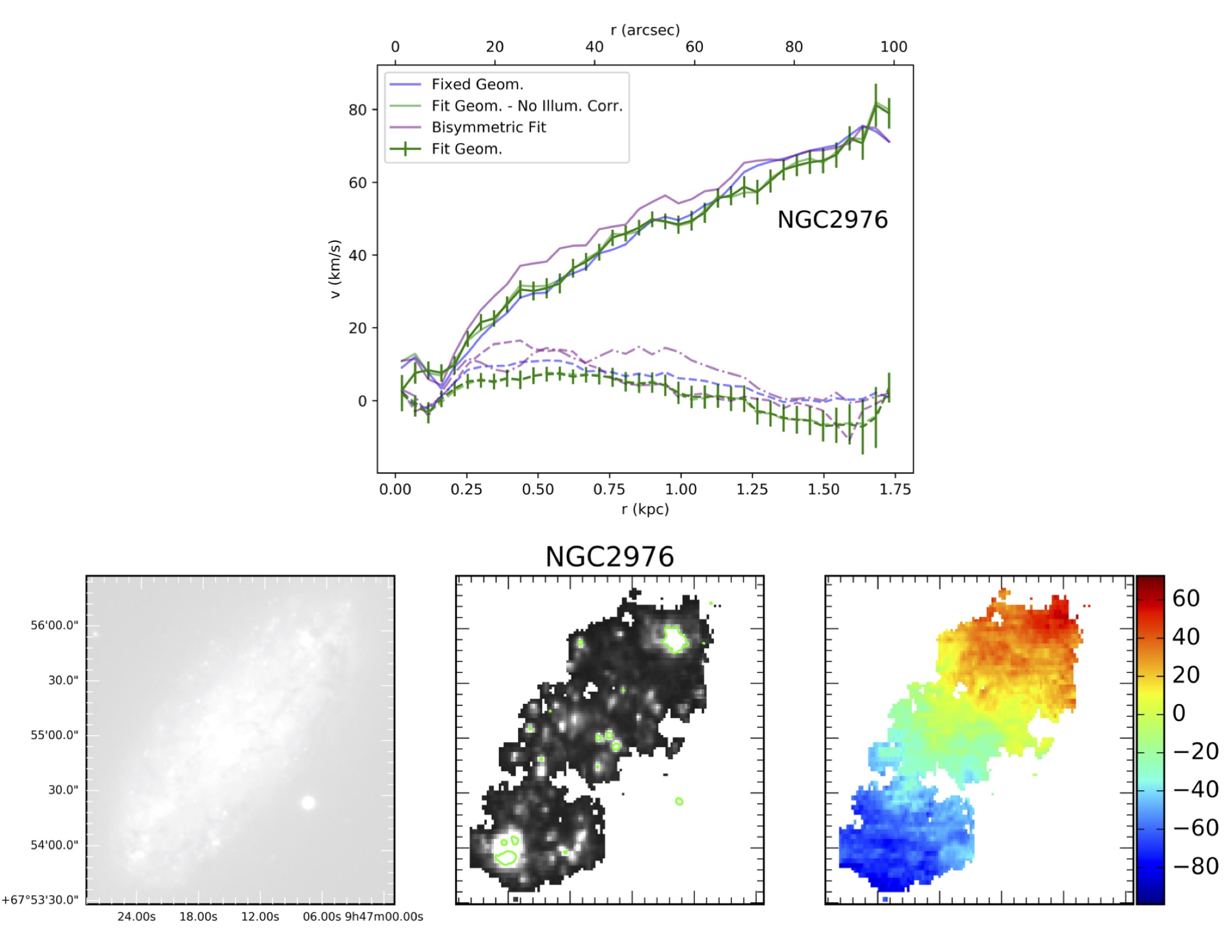}
\end{figure*}
\begin{figure*}
\centering
\includegraphics[width=0.83\textwidth]{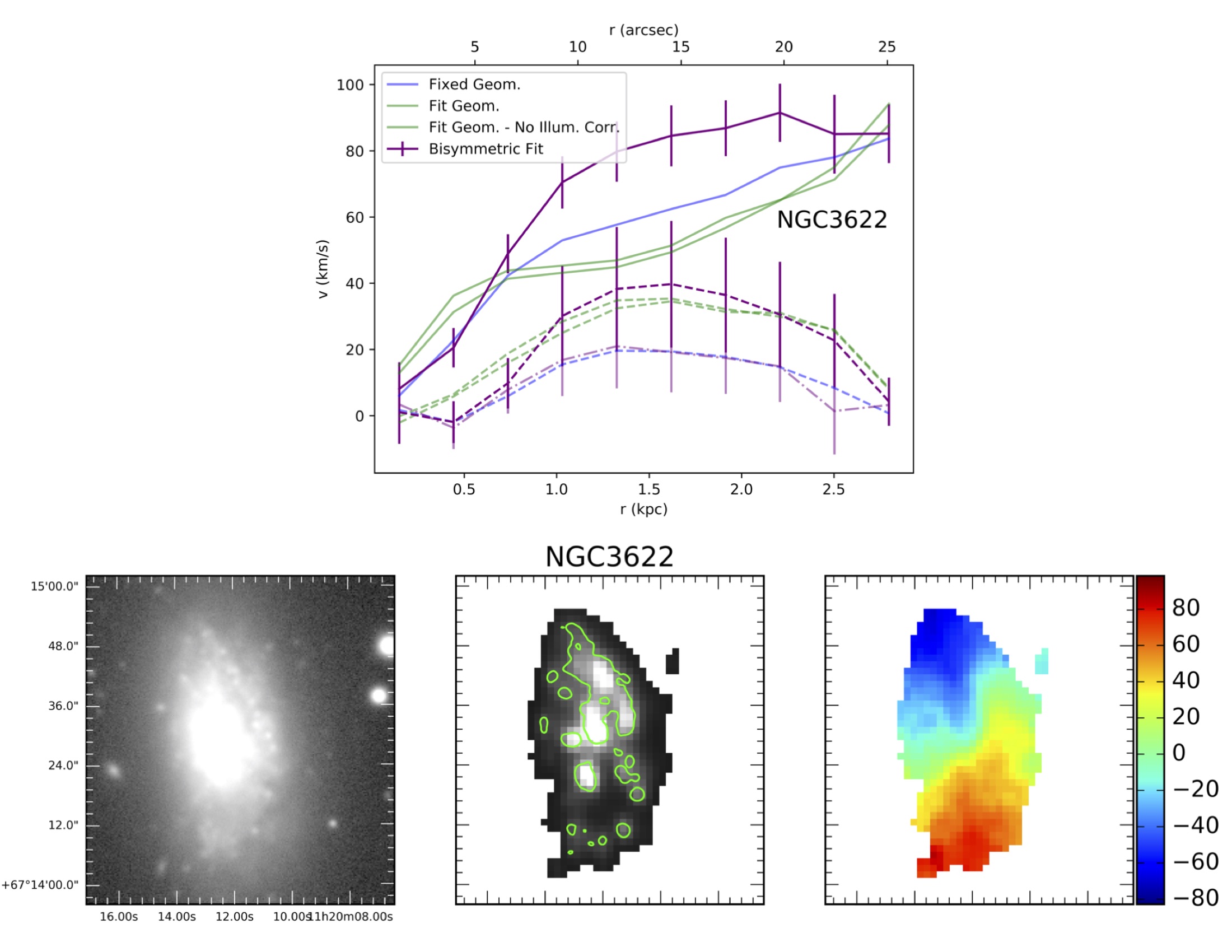}
\includegraphics[width=0.83\textwidth]{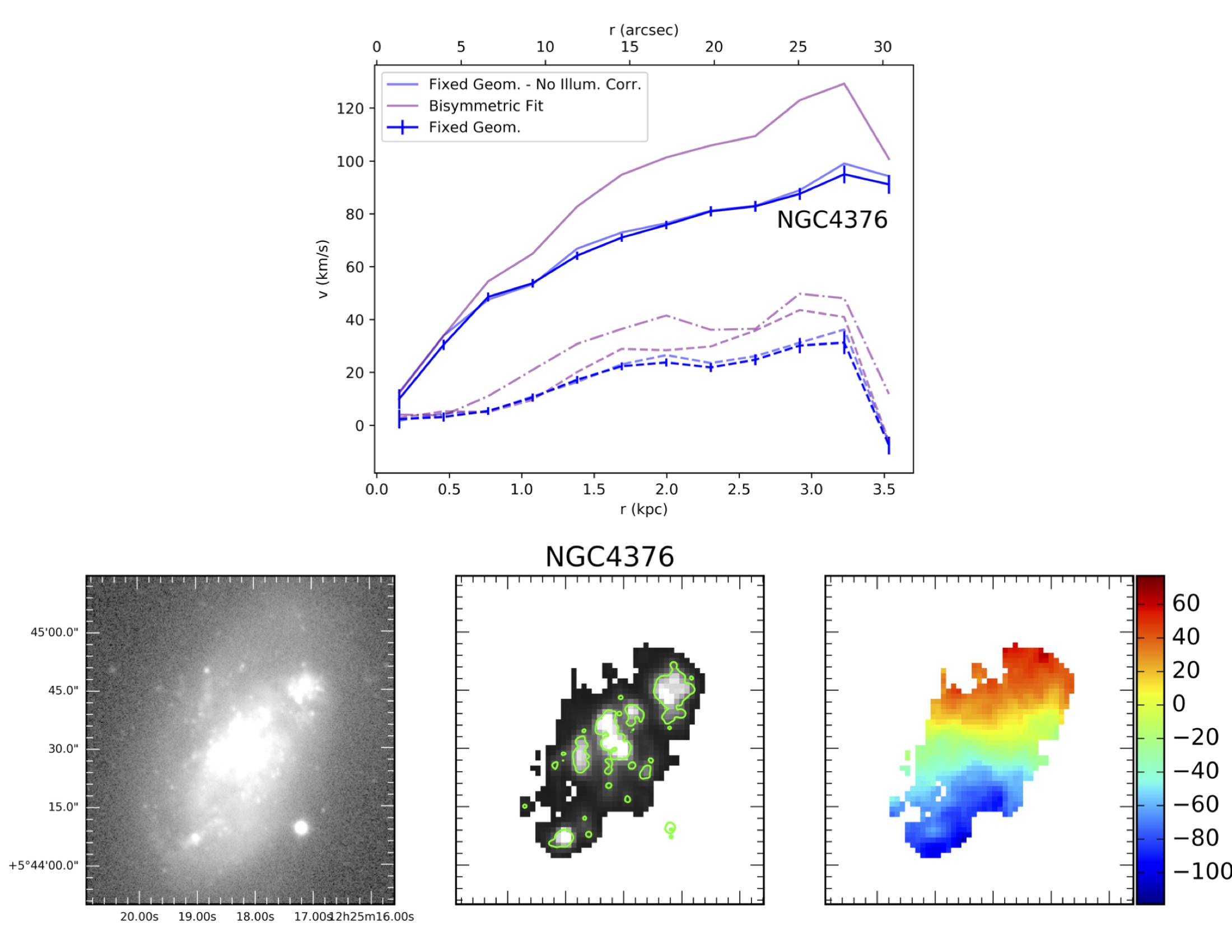}
\end{figure*}
\begin{figure*}
\centering
\includegraphics[width=0.83\textwidth]{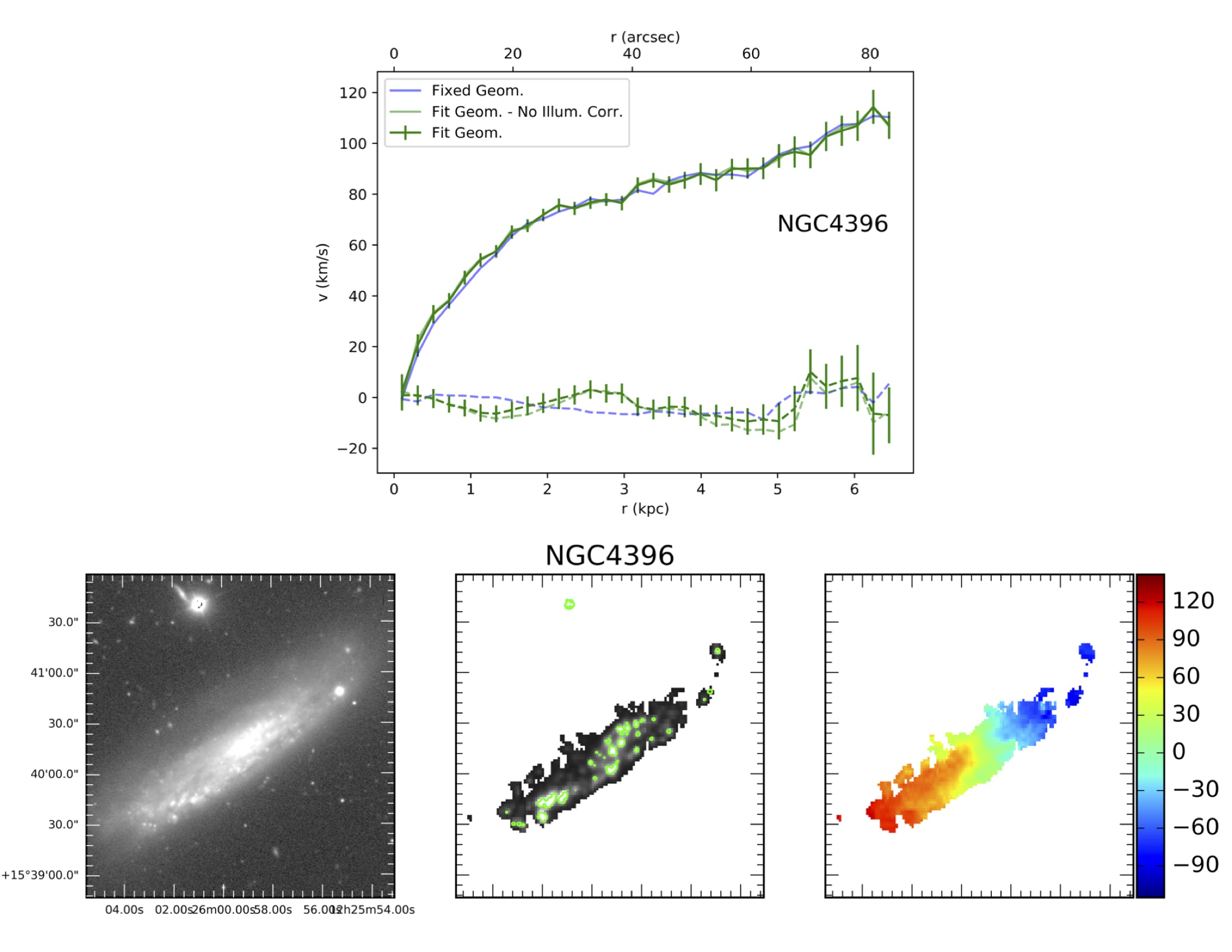}
\includegraphics[width=0.83\textwidth]{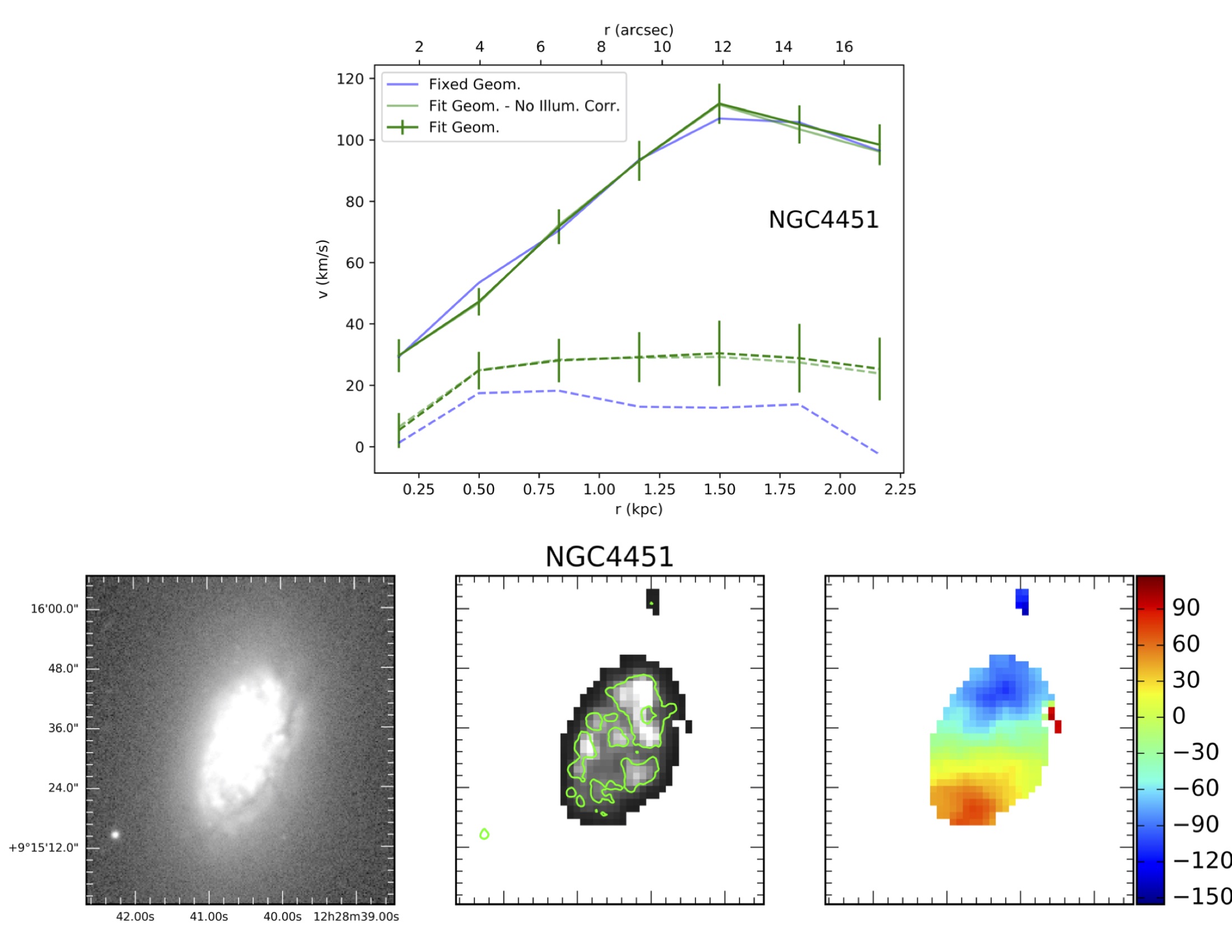}
\caption{Galaxies NGC 2644, NGC 2976, NGC 3622, NGC 4376, NGC 4396, and NGC 4451. For each, the top panel shows the H$\alpha$ rotation curve, while the bottom panels (from left to right) show the $r$-band image, total H$\alpha$ flux (with contours from photometry), and velocity field.}
\end{figure*}

\begin{figure*}
\centering
\includegraphics[width=0.83\textwidth]{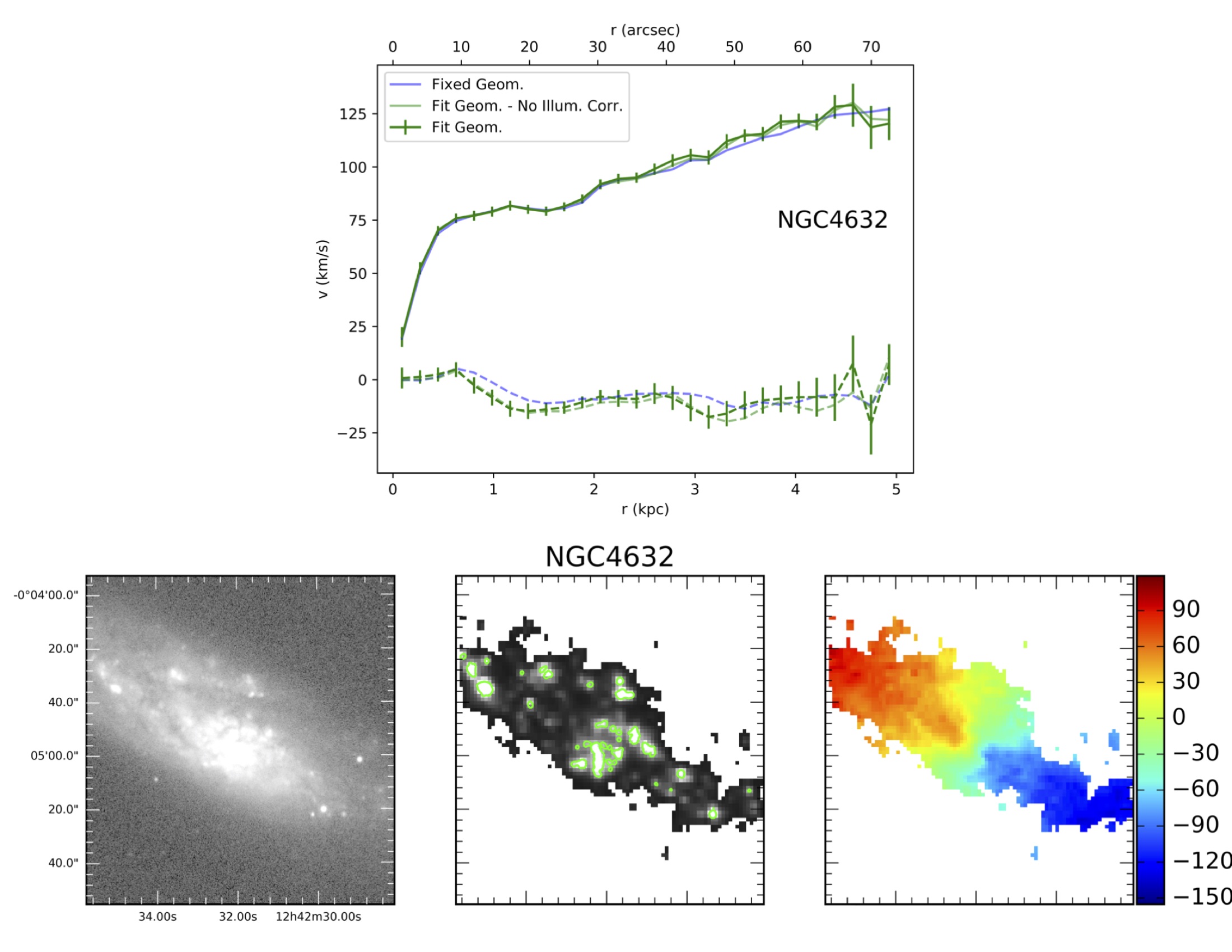}
\includegraphics[width=0.83\textwidth]{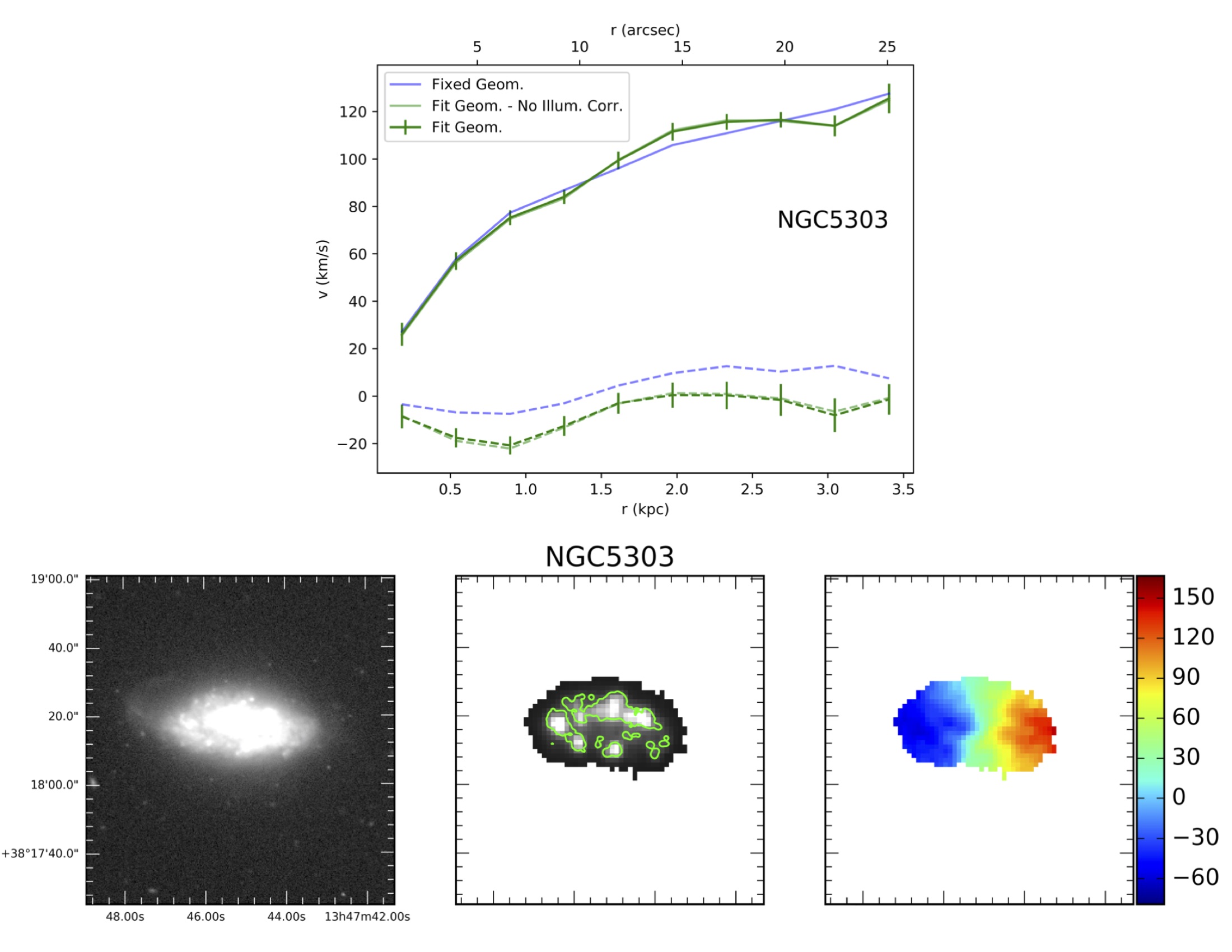}
\end{figure*}
\begin{figure*}
\centering
\includegraphics[width=0.83\textwidth]{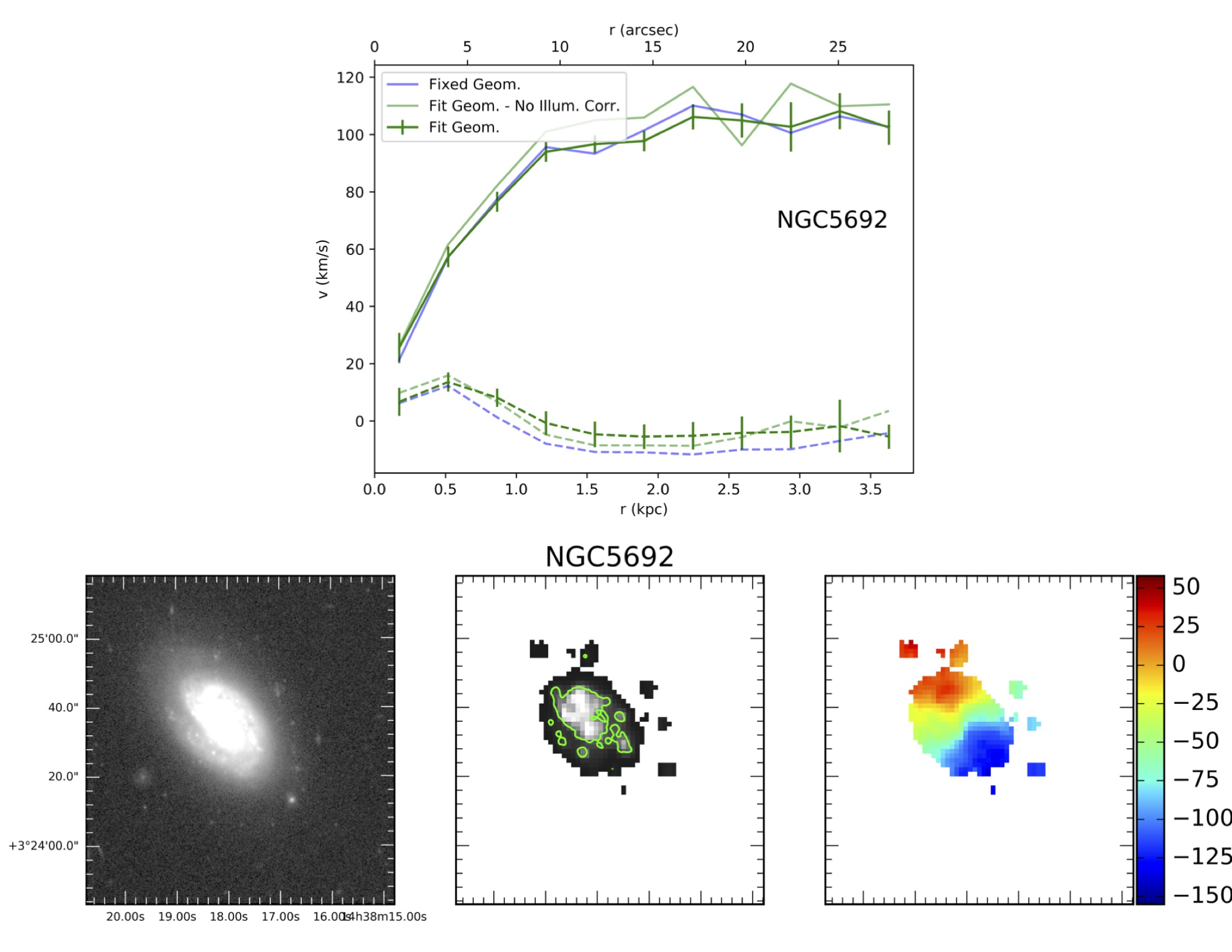}
\includegraphics[width=0.83\textwidth]{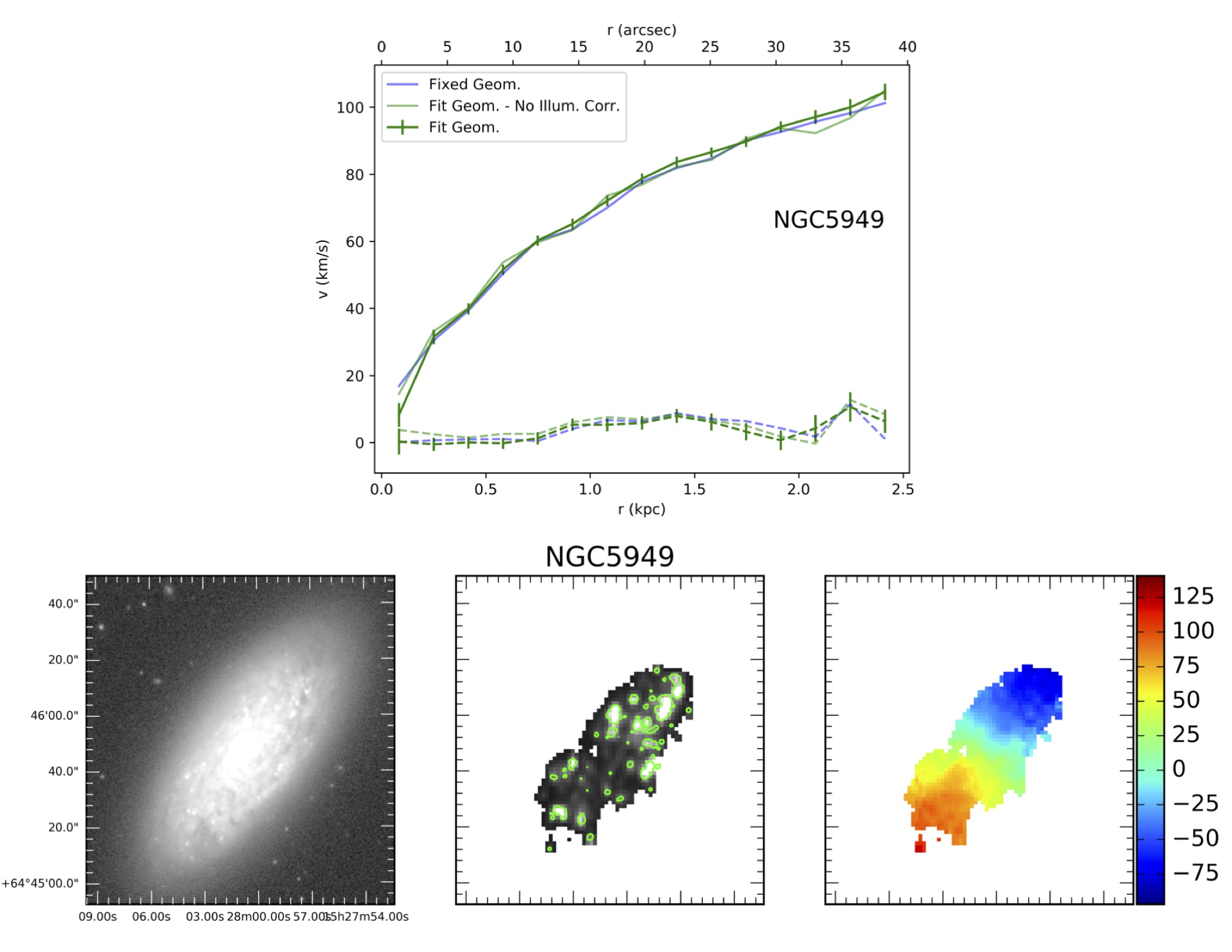}
\end{figure*}
\begin{figure*}
\centering
\includegraphics[width=0.83\textwidth]{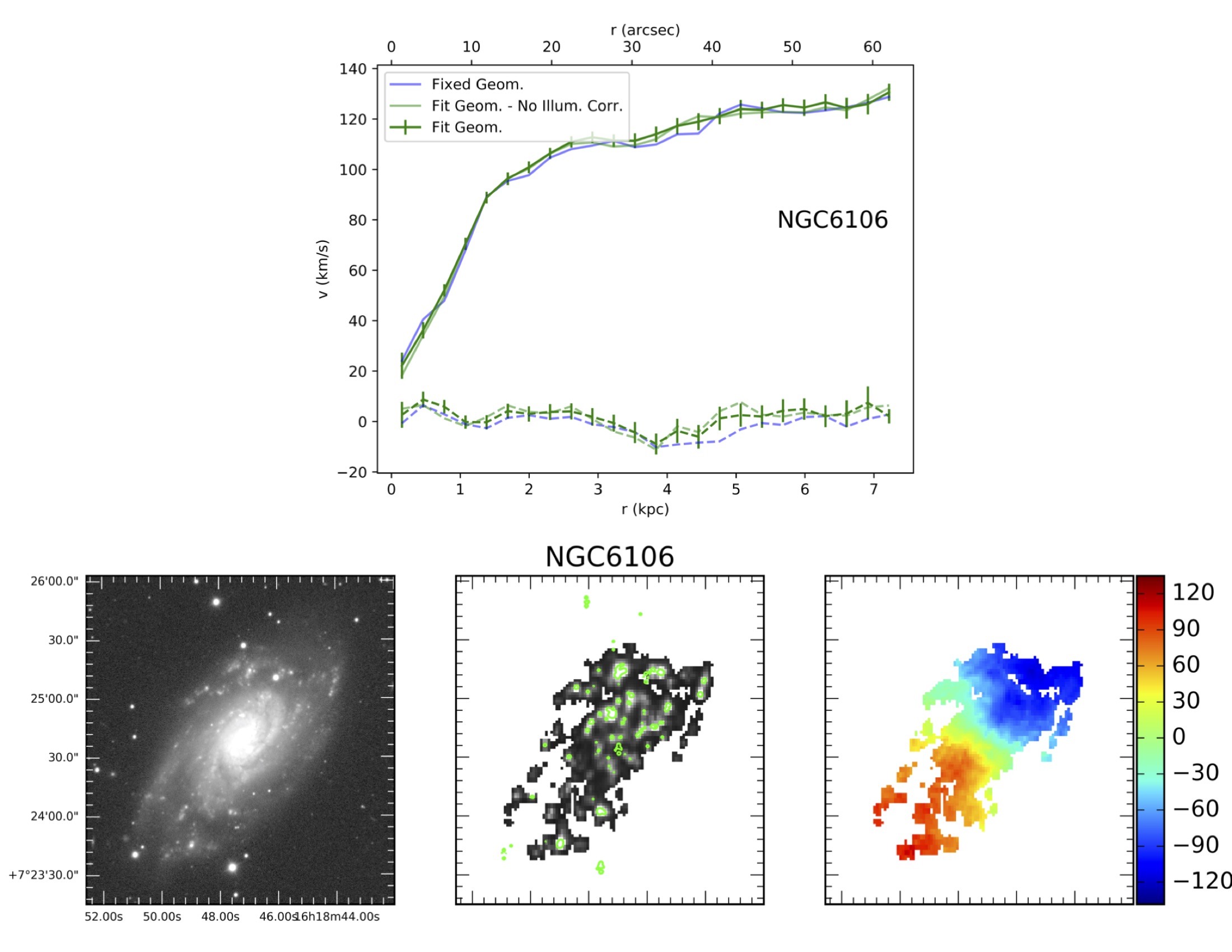}
\includegraphics[width=0.83\textwidth]{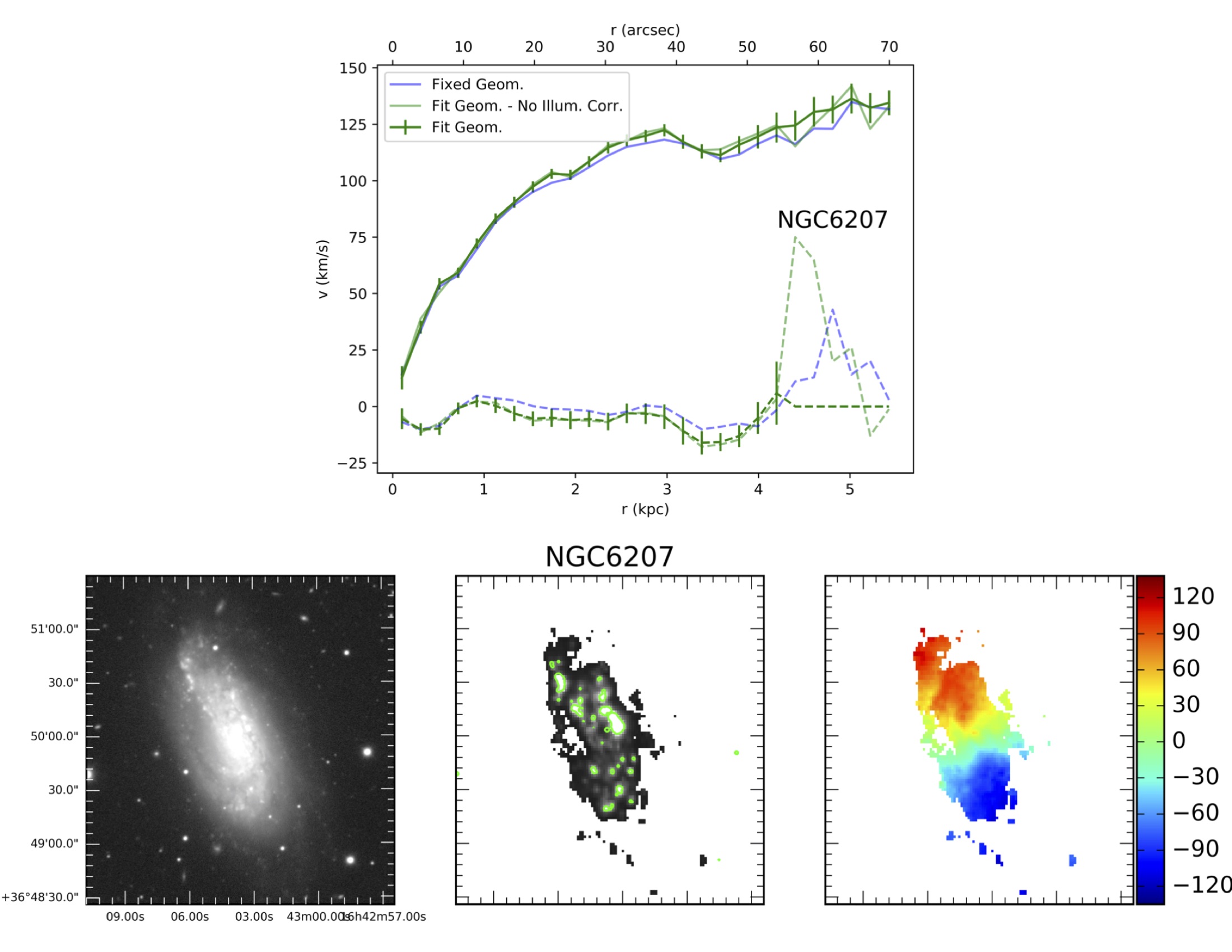}
\caption{Galaxies NGC 4632, NGC 5303, NGC 5692, NGC 5949, NGC 6106, and NGC 6207. For each, the top panel shows the H$\alpha$ rotation curve, while the bottom panels (from left to right) shows the $r$-band image, total H$\alpha$ flux (with contours from photometry), and velocity field.}
\end{figure*}

\begin{figure*}
\centering
\includegraphics[width=0.83\textwidth]{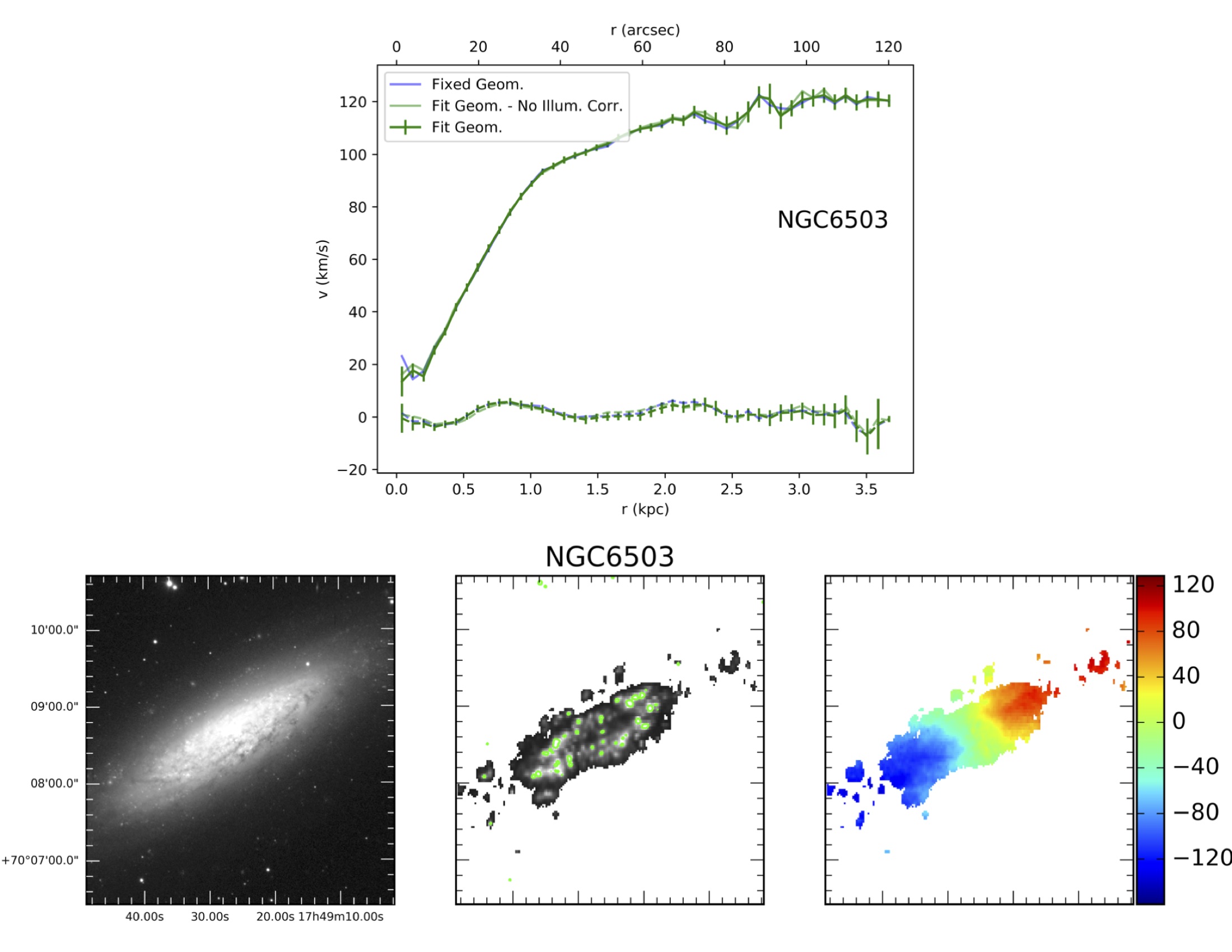}
\includegraphics[width=0.83\textwidth]{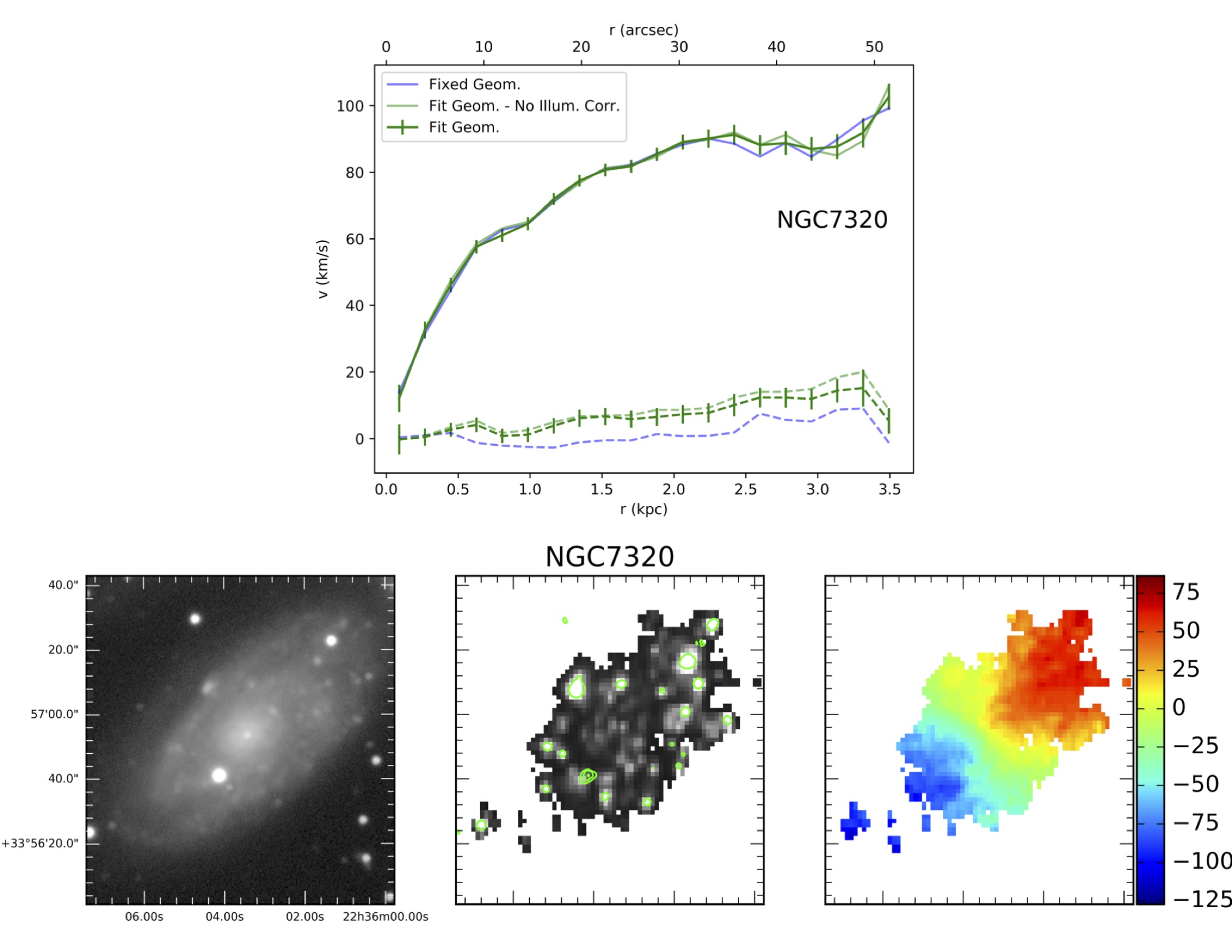}
\end{figure*}
\begin{figure*}
\centering
\includegraphics[width=0.83\textwidth]{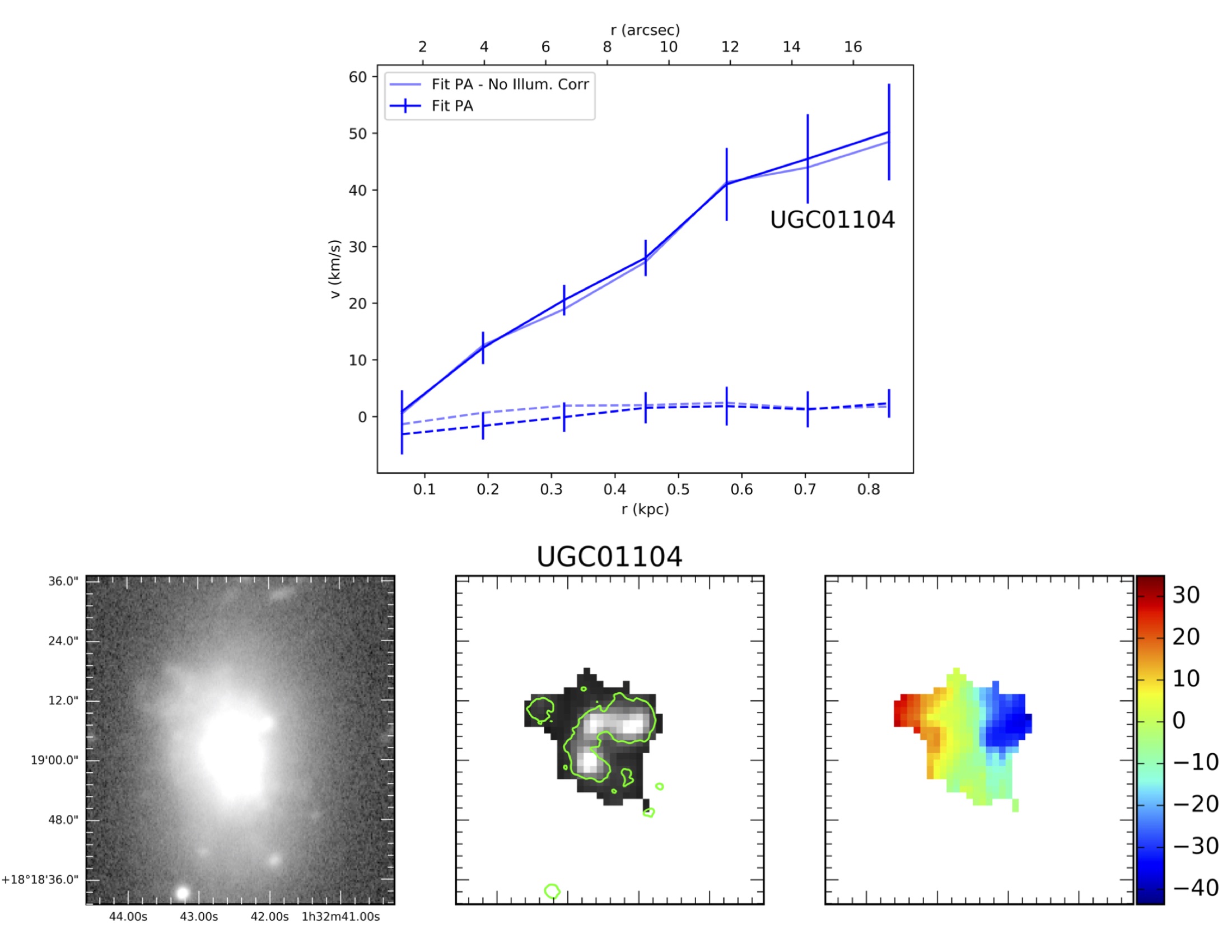}
\includegraphics[width=0.83\textwidth]{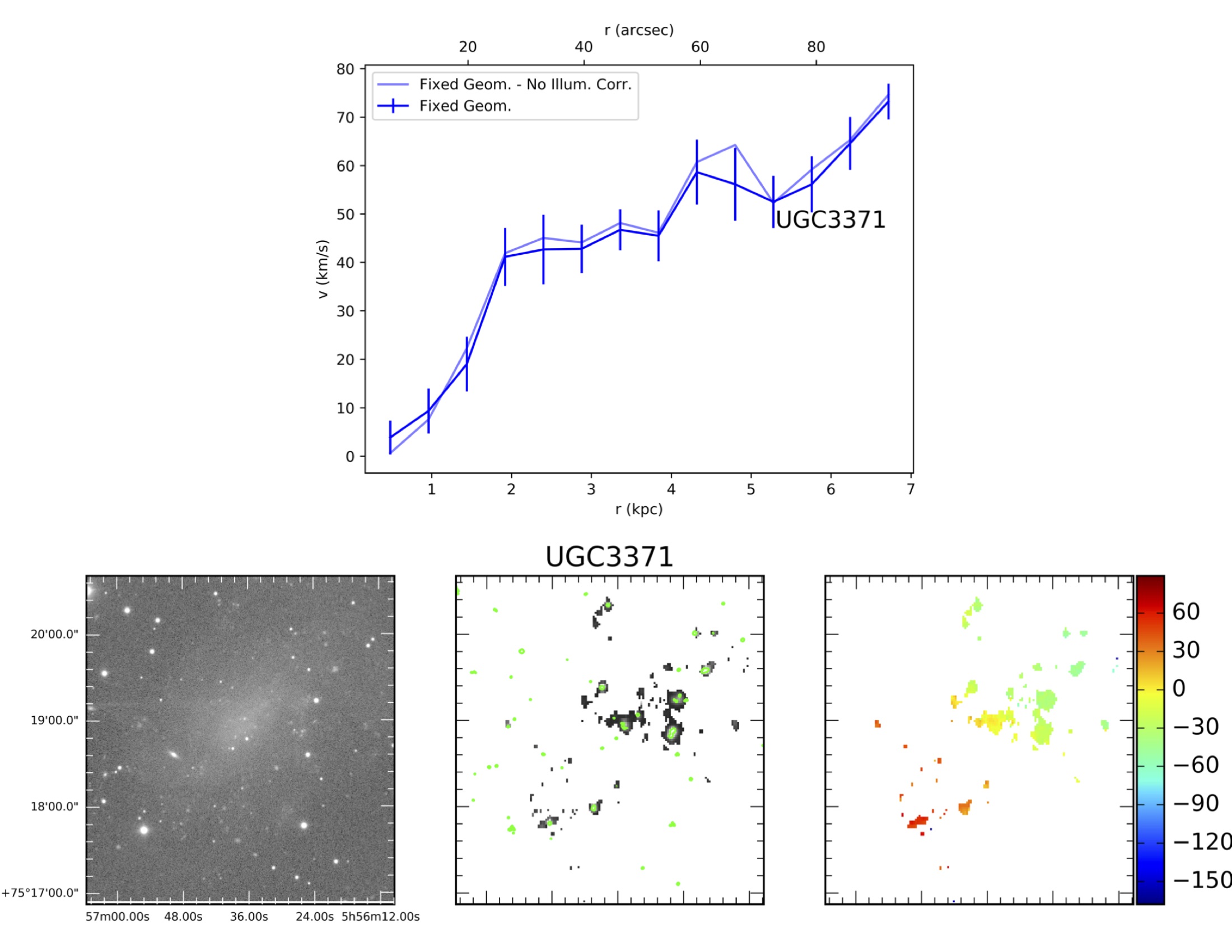}
\end{figure*}
\begin{figure*}
\centering
\includegraphics[width=0.83\textwidth]{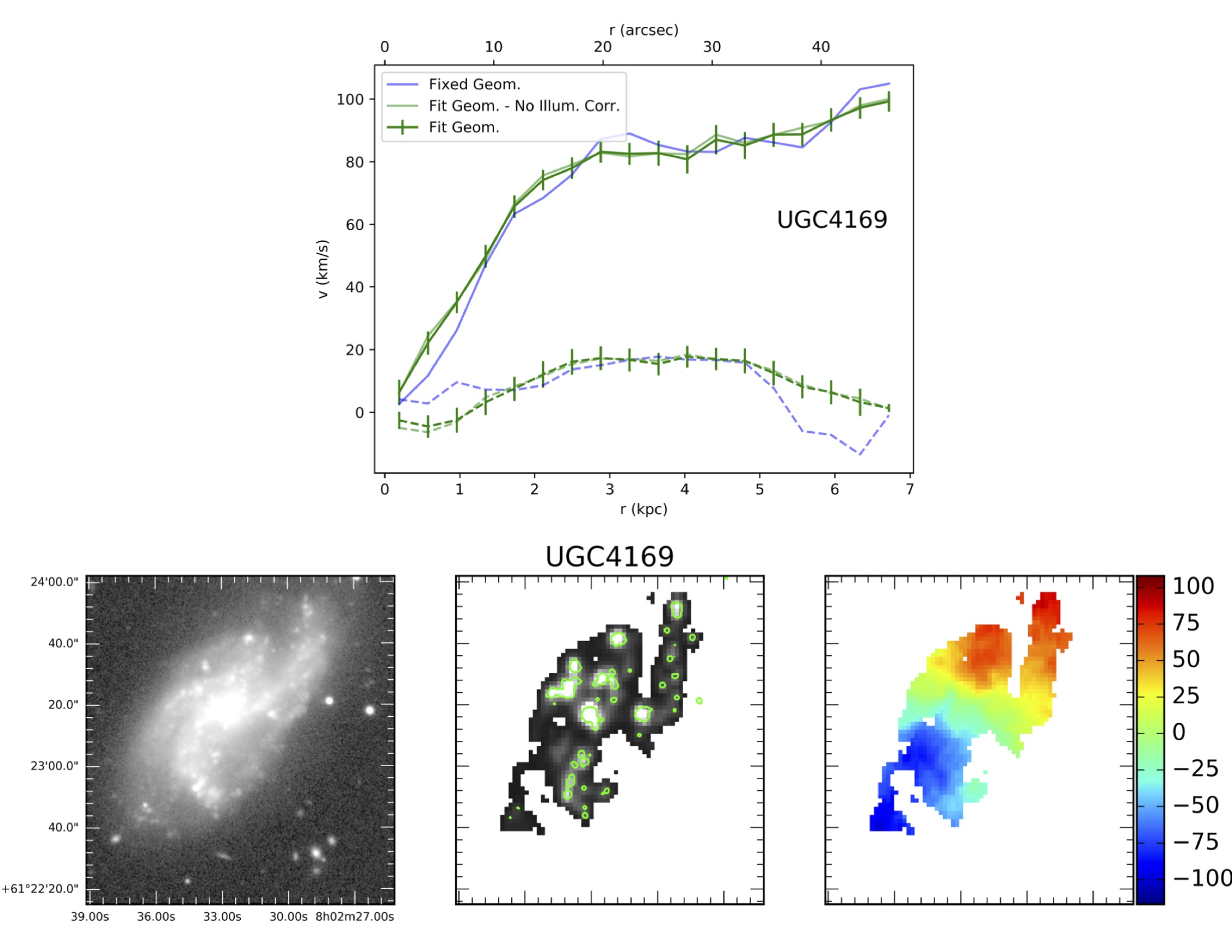}
\includegraphics[width=0.83\textwidth]{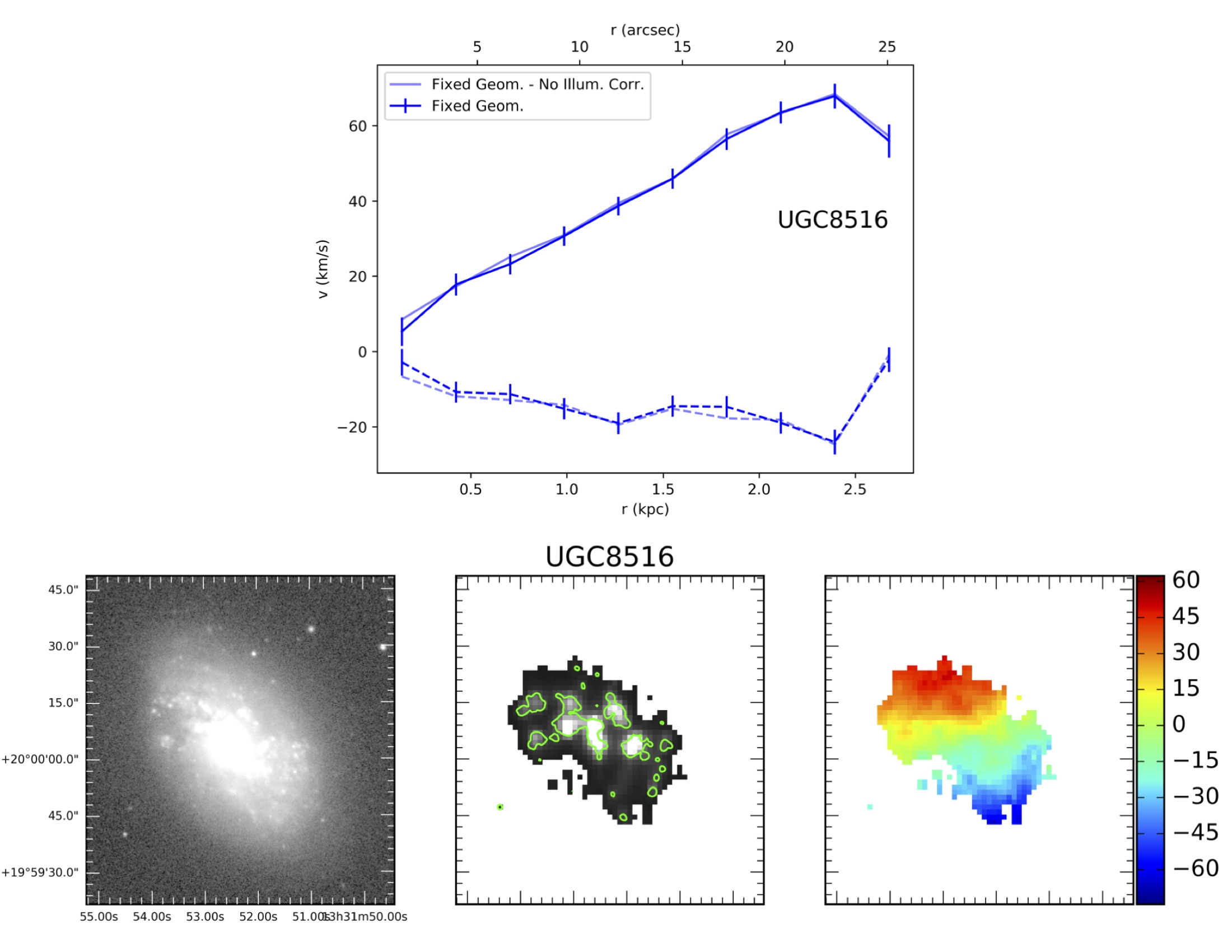}
\caption{Galaxies NGC 6503, NGC 7320, UGC 01104, UCG 3371, UGC 4169, and UGC 8516. For each, the top panel shows the H$\alpha$ rotation curve, while the bottom panels (from left to right) show the $r$-band image, total H$\alpha$ flux (with contours from photometry), and velocity field.}
\end{figure*}

\begin{figure*}
\centering
\includegraphics[width=0.83\textwidth]{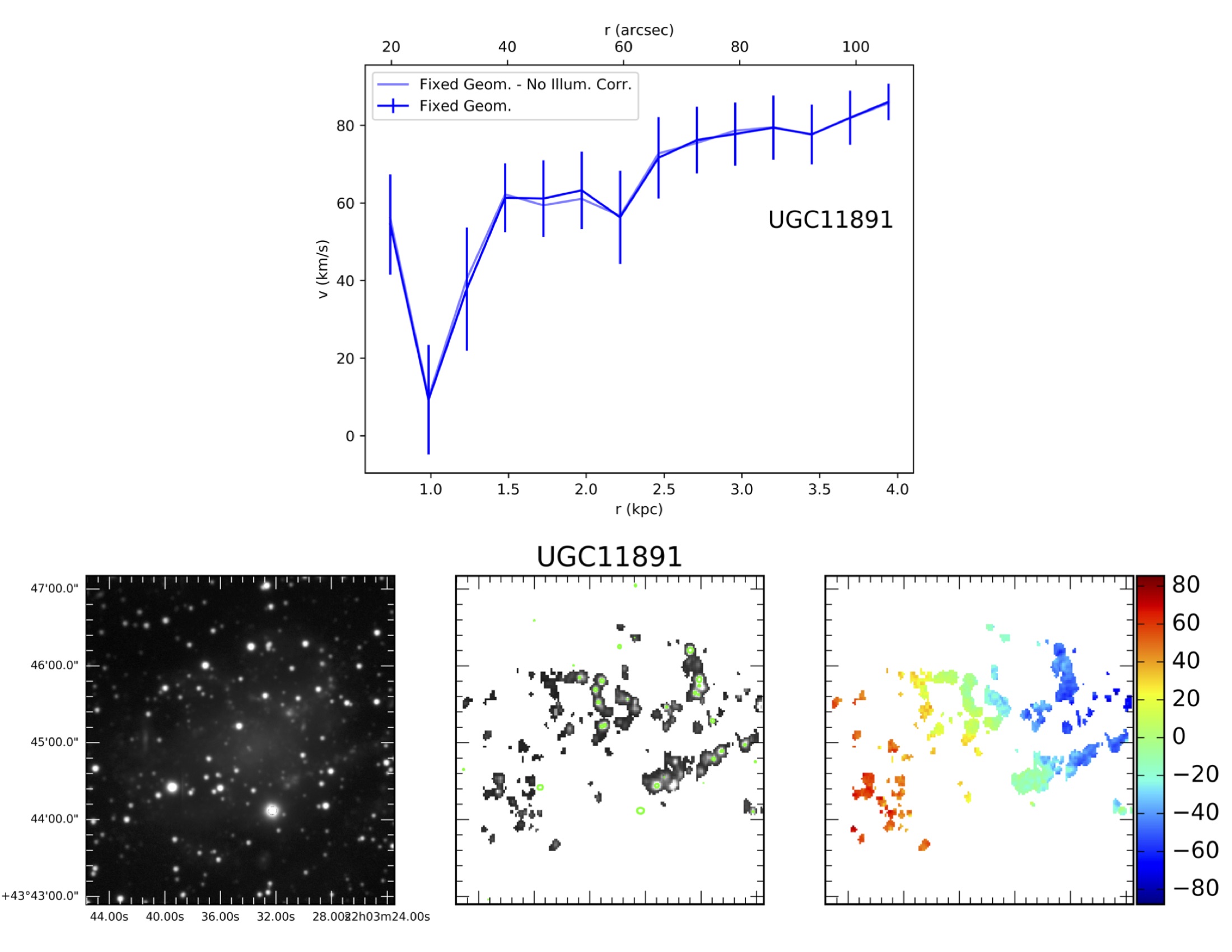}
\includegraphics[width=0.83\textwidth]{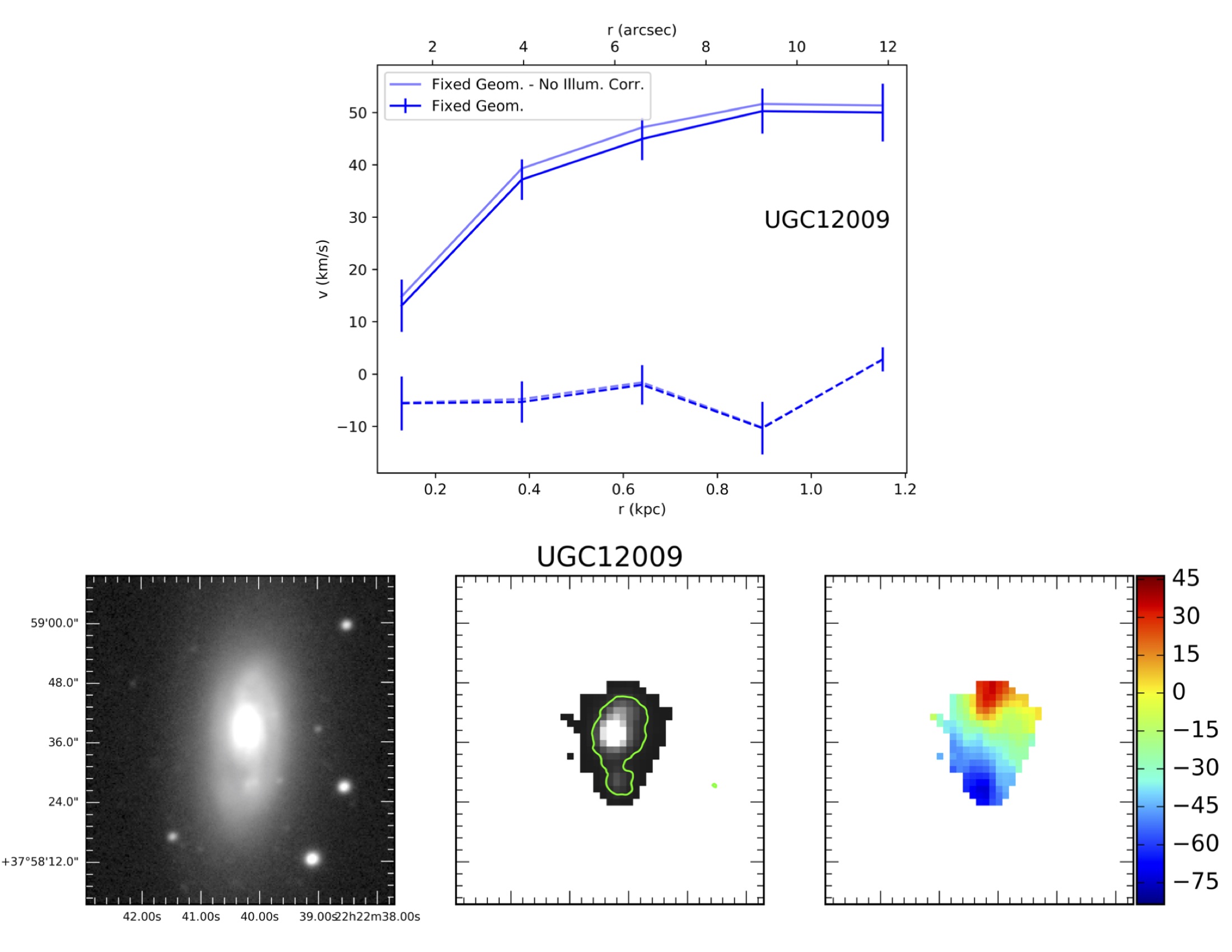}
\caption{Galaxies UGC 11891 and UGC 12009. For each, the top panel shows the H$\alpha$ rotation curve, while the bottom panels (from left to right) show the $r$-band image, total H$\alpha$ flux (with contours from photometry), and velocity field.}
\end{figure*}

\bibliographystyle{apj}
\bibliography{Relatores}

\end{document}